\documentclass[10pt]{article}
\usepackage{fullpage}
\usepackage{amsmath}
\usepackage{amssymb}
\usepackage[dvips]{epsfig}

\usepackage{amssymb}
\usepackage{amsmath}
\usepackage{epsf}
\usepackage{graphicx}
\usepackage[dvips]{epsfig}
\usepackage{bm}
\usepackage{color}

\def\##1{{\bf #1}}
\def\=#1{\underline{\underline{#1}}}

\def\+#1{\underline{\bf #1}}
\def\*#1{\underline{\underline{\bf #1}}}

\def\r#1{(\ref{#1})}
\def\l#1{\label{#1}}
\def\c#1{\cite{#1}}

\def\le{\left(}
\def\ri{\right)}
\def\les{\left[}
\def\ris{\right]}
\def\lec{\left\{}
\def\ric{\right\}}

\def\.{\mbox{ \tiny{$^\bullet$} }}

\def\epso{\epsilon_{\scriptscriptstyle 0}}
\def\lambdao{\lambda_{\scriptscriptstyle 0}}
\def\muo{\mu_{\scriptscriptstyle 0}}

\def\ko{k_{\scriptscriptstyle 0}}
\def\wo{w_{\scriptscriptstyle 0}}
\def\co{c_{\scriptscriptstyle 0}}

\def\eps{\epsilon}

\begin{document}

\begin{center}

{\bf {\LARGE Counterposition and
 negative phase velocity \vspace{4pt} \\
 in
 uniformly moving dissipative materials}}

\vspace{10mm} \large

 Tom G. Mackay\footnote{E--mail: T.Mackay@ed.ac.uk.}\\
{\em School of Mathematics and
   Maxwell Institute for Mathematical Sciences\\
University of Edinburgh, Edinburgh EH9 3JZ, UK}\\
and\\
 {\em NanoMM~---~Nanoengineered Metamaterials Group\\ Department of Engineering Science and Mechanics\\
Pennsylvania State University, University Park, PA 16802--6812,
USA}\\
 \vspace{3mm}
 Akhlesh  Lakhtakia\footnote{E--mail: akhlesh@psu.edu}\\
 {\em NanoMM~---~Nanoengineered Metamaterials Group\\ Department of Engineering Science and Mechanics\\
Pennsylvania State University, University Park, PA 16802--6812, USA}

\end{center}

\vspace{4mm}

\begin{abstract}

The Lorentz transformations of electric and magnetic fields were
implemented to study (i) the refraction of linearly polarized plane
waves into a half--space occupied by a uniformly moving material,
 and (ii) the traversal of
linearly polarized Gaussian beams through a uniformly moving slab.
Motion was taken to occur tangentially to the interface(s) and in
the plane of incidence. The moving materials were assumed to be
isotropic, homogeneous, dissipative dielectric materials from the
perspective of a co--moving observer.
 Two different  moving materials were considered: from the perspective of a co--moving observer,
  material A
 supports planewave propagation with only positive phase velocity,
whereas material B  supports planewave propagation with both
positive and negative phase velocity, depending on the polarization
state. For both materials A and B, the sense of the phase velocity
and whether or not counterposition occurred, as perceived by a
nonco--moving observer, could be altered by varying the observer's
velocity. Furthermore, the  lateral position of a beam upon
propagating through a uniformly moving slab made of material A, as
perceived by a nonco--moving observer, could be controlled by
varying the observer's velocity. In particular, at certain
velocities, the transmitted beam  emerged from the slab laterally
displaced in the direction opposite to the direction of incident
beam. The transmittances of a uniformly moving slab made of material
B  were very small and the energy density of the transmitted beam
was largely concentrated in the direction normal to the slab,
regardless of the observer's velocity.

\end{abstract}



\section{Introduction}

What happens to plane waves at the planar interface of two passive
linear mediums  is one of the staple problems of electrodynamics.
Relative motion between the two mediums on either side of the
interface can give rise to  complex electromagnetic
behaviour~---~exemplified by
  materials which
support negative refraction \c{SAR}~---~even when both mediums are
simple isotropic  dielectric--magnetic mediums with respect to a
co--moving observer.

 In this paper we chiefly focus on the effects of relative uniform motion on the
 phenomenons of
\emph{counterposition} and \emph{negative phase velocity} (NPV). The
orientation of the cycle--averaged Poynting vector is central to
both of  these phenomenons.
 The phase velocity is said to be  negative (positive) if it casts
a negative (positive) projection onto the the cycle--averaged
Poynting vector. NPV has commonly been understood to betoken
negative refraction  \c{Laeu,EJP}.
 However, certain anisotropic mediums can
support
 NPV propagation in conjunction with positive refraction (and vice
versa) \c{Belov_MOTL}, and
 we recently demonstrated that the correspondence between NPV and
 negative refraction is only strictly appropriate in the case of
uniform planewave propagation in isotropic dielectric--magnetic
mediums \c{ML_PRB}. Usually, at a planar interface,  the
cycle--averaged Poynting vector and the real part of the wavevector
associated with the refracted planewave are oriented on the same
side of the normal to the planar interface \c{Chen}. But under
certain  conditions, these two vectors can be counterposed, i.e.,
oriented on opposite sides of the normal to the planar interface.
This counterposition has been described in certain isotropic
dielectric \c{ML_PRB} and anisotropic mediums
\c{Belov_MOTL,ZFM,Optik_counterposition} in nonrelativistic
scenarios, as well as in certain relativistic scenarios
\c{Kong_PRB,MOTL_counterposition,note1}.

Previous theoretical studies concerning counterposition and NPV in
uniformly moving mediums have mostly relied  upon the Minkowski
constitutive relations to describe the moving medium  in a
nonco--moving inertial reference frame
\c{Kong_PRB,MOTL_counterposition,Hiding}, following the standard
textbook approach \c{Chen}. These studies have shown that
nondissipative mediums which do not support NPV and counterposition
from the perspective of a co--moving observer may support NPV from
the perspective of certain nonco--moving inertial reference frames.
Furthermore, by considering the propagation of  a Gaussian beam
through a uniformly moving, nondissipative slab, we found that a
degree of concealment could be achieved under certain conditions
\c{Hiding}. However, the Minkowski constitutive relations are
strictly appropriate only for instantaneously responding mediums
\c{BC}. For realistic material mediums, it is necessary to
incorporate both spatial and temporal nonlocality in the
nonco--moving inertial reference frame.

Recently, using a direct approach in which the Lorentz
transformations of the electric and magnetic fields are implemented,
 a dissipative medium which does not support NPV from the
perspective of a co--moving observer was observed to support NPV
from the perspective of certain nonco--moving inertial reference
frames \c{ML_Optik}. This direct approach~---~which does not involve
the Minkowski constitutive relations~---~is implemented in the
following sections to investigate whether counterposition and NPV
can be induced in uniformly moving, realistic materials and, if so,
under what circumstances. A particular focus of our attention is the
behaviour of an isotropic dielectric material which supports the
propagation of nonuniform plane waves with both positive and
negative phase velocity, depending upon the polarization state,  in
a co--moving inertial reference frame. This unexpected property for
an isotropic dielectric material came to light only very recently
\c{ML_PRB}.

As regards notation, 3--vectors are in boldface with the $\hat{}$
symbol denoting a unit vector; and the identity 3$\times$3  dyadic
is expressed as $\=I = \hat{\#x}\,\hat{\#x} +  \hat{\#y}\,\hat{\#y}
+ \hat{\#z}\,\hat{\#z}$. The operators Re and Im yield the real and
imaginary parts of complex quantities; and $i = \sqrt{-1}$. The
permittivity and permeability of free space are written as $\epso$
and $\muo$, respectively; and $\co = 1 / \sqrt{ \epso \muo}$ is the
speed of light in free space.

\section{Refraction into a moving half--space}
\l{moving_hs}

\subsection{Theoretical considerations}

Let us  consider a homogeneous material which is a  spatially local,
isotropic,  dielectric material,  characterized by the
 frequency--domain constitutive
 relations
\begin{equation}
\left.
\begin{array}{l}
\#D' =   \epso \eps'\#E'  \\ \vspace{-3mm} \\
\#B' =  \muo \#H'
\end{array}
\right\}, \l{Con_rel_p}
\end{equation}
in an inertial reference  frame  $\Sigma'$. Since the material is
generally dissipative, the
  relative permittivity $\eps'$
 is a complex--valued function of the angular frequency
 $\omega'$. This material occupies the half--space $z>0$, whereas the
 half--space $z< 0$ is vacuous.

 The  inertial reference frame $\Sigma'$ moves at constant velocity
$\#v=v\hat{\#v}$ relative to the inertial reference frame
 $\Sigma$. We choose $\hat{\#v} = \hat{\#x}$.
 The electromagnetic field phasors in $\Sigma$ are related to those
in $\Sigma'$ by the Lorentz transformations \c{Chen}
\begin{equation}
\left. \begin{array}{l} \#E = \displaystyle{\le \#E' \.\hat{\#v} \ri
\hat{\#v} +
 \gamma \,
\les \le \=I - \hat{\#v}\hat{\#v} \ri \. \#E' - \#v \times
\#B' \ris } \vspace{4pt}  \\
\#B = \displaystyle{\le \#B' \.\hat{\#v} \ri \hat{\#v} +
 \gamma \,
\les \le \=I - \hat{\#v}\hat{\#v} \ri \. \#B' + \frac{ \#v \times
\#E'}{\co^2} \ris  } \vspace{4pt} \\
\#H = \displaystyle{\le \#H' \.\hat{\#v} \ri \hat{\#v} +
 \gamma \,
\les \le \=I - \hat{\#v}\hat{\#v} \ri \. \#H' + \#v \times
\#D' \ris  } \vspace{4pt} \\
\#D = \displaystyle{\le \#D' \.\hat{\#v} \ri \hat{\#v} +
 \gamma \,
\les \le \=I - \hat{\#v}\hat{\#v} \ri \. \#D' - \frac{ \#v \times
\#H'}{\co^2} \ris }\end{array} \right\},  \l{Dp}
\end{equation}
where the scalars
\begin{equation}
\gamma = \frac{1}{\sqrt{1 - \beta^2}},\qquad \beta =
\frac{v}{\co}\,.
\end{equation}

An infinitely long line source oriented parallel to the $y$ axis
 is located in the vacuous half--space $z<0$, far from the interface
$z=0$. The source is stationary in the inertial reference
frame $\Sigma$. The field launched by it can be decomposed into an
angular spectrum of plane waves. Therefore, it suffices to consider
just one plane wave  incident on the interface $z=0$.

This plane
wave is described in $\Sigma$  by the electric and magnetic field
phasors
\begin{equation}
\left.\begin{array}{l}
\#E_i  = \#e_{i}\, \exp \les i \le \#k_{i}\cdot\#r - \omega t \ri \ris   \\[5pt]
\#H_i  = \#h_{i} \, \exp \les i \le \#k_{i} \cdot\#r   - \omega
t \ri \ris
\end{array}\right\}, \qquad z \leq 0,
\l{pw_vac}
\end{equation}
 where the wavevector
\begin{equation}
\#k_{i} = \ko \le \sin \theta \, \hat{\#x} + \cos \theta \,
\hat{\#z} \ri,
\end{equation}
with the free--space wavenumber  $ \ko = \omega \sqrt{\epso \muo}$
and $\omega$ being the angular frequency with respect to $\Sigma$.
As the incident plane wave transports energy towards the interface,
the angle $\theta\in\le -90^\circ, \, 90^\circ\ri$ and hence the
real--valued scalar
\begin{equation}
\kappa = \ko\sin\theta \in \left( -\ko, \, \ko \right)\,.
\end{equation}
Thus, we have chosen the velocity of the moving material as
tangential to the interface plane $z=0$ and lying in the plane of
incidence (i.e., the $xz$ plane).

 In $\Sigma'$, the counterparts of the phasors \r{pw_vac} are
\begin{equation}
\left.\begin{array}{l}
\#E'_i  = \#e'_{i}\, \exp \les i \le \#k'_{i}\cdot\#r' - \omega' t' \ri \ris   \\[5pt]
\#H'_i = \#h'_{i} \, \exp \les i \le \#k'_{i} \cdot\#r'   -
\omega' t' \ri \ris
\end{array}\right\},\qquad z \leq 0,
\l{pw_vac_p}
\end{equation}
wherein the  phasor amplitudes $\lec \#e'_i, \#h'_i \ric$
are related to $\lec \#e_i, \#h_i
\ric$
  via the Lorentz transformations \r{Dp}, while
\begin{equation}
\left.
\begin{array}{l}
 \#k_{i} = \displaystyle{ \gamma \le \#k'_{i} \. \hat{\#v} + \frac{\omega' v}{\co^2}
\ri
\hat{\#v} + \le \, \=I - \hat{\#v} \, \hat{\#v} \ri \. \#k'_{i}} \vspace{6pt}\\
\#r = \displaystyle{\les \, \=I  + \le \gamma - 1 \ri
\hat{\#v}\,\hat{\#v} \ris \.
\#r' + \gamma \, \#v t'}\vspace{6pt} \\
 \omega = \displaystyle{\gamma \le \omega' + \#k'_{i} \. \#v \ri} \vspace{6pt}\\
 t = \displaystyle{\gamma \le t' + \frac{\#v \. \#r'}{\co^2} \ri}
 \end{array} \right\}. \l{wT}
 \end{equation}

The  plane wave  refracted into the half--space $z
> 0$ is represented in $\Sigma'$ by the electric and magnetic field phasors
\begin{equation}
\left.\begin{array}{l}
\#E'_t  = \#e'_{t}\, \exp \les  i \le \#k'_{t}\cdot\#r' - \omega' t' \ri \ris  \\[5pt]
\#H'_t  = \#h'_{t} \, \exp\les i \le  \#k'_{t} \cdot\#r'   - \omega'
t' \ri \ris
\end{array}\right\}, \qquad z \geq 0,
\l{pw_mat}
\end{equation}
wherein the wavevector
\begin{equation}
\#k'_{t} =  \left(\#k'_i \.\hat{\#x}\right) \hat{\#x} + k'_{zt} \, \hat{\#z}
\end{equation}
adheres to Snel's law.
Use of the constitutive relations \r{Con_rel_p}
and application of the Maxwell curl postulates in $\Sigma'$ provides an expression
for $k'_{zt}$ and relationships
between $\#e'_{t}$ and $\#h'_{t}$ \c{Chen}.

With respect to $\Sigma'$, the reflected plane wave may be expressed
as
\begin{equation}
\left.\begin{array}{l}
\#E'_r  = \#e'_{r}\, \exp \les i \le \#k'_{r}\cdot\#r' - \omega' t' \ri \ris   \\[5pt]
\#H'_r = \#h'_{r} \, \exp \les i \le \#k'_{r} \cdot\#r'   -
\omega' t' \ri \ris
\end{array}\right\},\qquad z \leq 0,
\l{pwr_vac_p}
\end{equation}
where
\begin{equation}
\#k'_{r} =  \left(\#k'_i \.\hat{\#x}\right) \hat{\#x} + k'_{zr} \, \hat{\#z}\,.
\end{equation}
Application of the Maxwell curl postulates in $\Sigma'$ provides an expression
for $k'_{zr}$ and relationships
between $\#e'_{r}$ and $\#h'_{r}$.

At this point, the usual boundary conditions can be invoked across
the plane $z=0$, i.e.,
\begin{equation}
\left.\begin{array}{ll}
(\#e'_i+ \#e'_r)\cdot\hat{\#x}= \#e'_t\cdot\hat{\#x}\,
&
(\#e'_i+ \#e'_r)\cdot\hat{\#y}= \#e'_t\cdot\hat{\#y}\\[5pt]
(\#h'_i+ \#h'_r)\cdot\hat{\#x}= \#h'_t\cdot\hat{\#x}\,
&
(\#h'_i+ \#h'_r)\cdot\hat{\#y}= \#h'_t\cdot\hat{\#y}
\end{array}\right\},\qquad z =0\,.
\end{equation}
In this way, the  phasor amplitudes $\lec \#e'_r, \#h'_r \ric$ and
$\lec \#e'_t, \#h'_t \ric$ can be determined. Thereafter, the
reflected and the refracted plane waves can be transformed back to
$\Sigma$. Most importantly in the present context,
 the refracted  plane wave turns out to be represented in $\Sigma$ by the electric and magnetic field phasors
\begin{equation}
\left.\begin{array}{l}
\#E_t  = \#e_{t}\, \exp \les  i \le \#k_{t}\cdot\#r - \omega t \ri \ris  \\[5pt]
\#H_t  = \#h_{t} \, \exp\les i \le  \#k_{t} \cdot\#r   - \omega
t \ri \ris
\end{array}\right\}, \qquad z \geq 0,
\l{pw_mat2}
\end{equation}
wherein the wavevector
\begin{equation}
\#k_{t} =   \kappa \hat{\#x} + k_{zt} \, \hat{\#z}.
\end{equation}
 Since  $k_{zt} \in \mathbb{C}$, the refracted plane wave is
 generally  nonuniform.

\subsection{Numerical studies} \l{num_1}

\begin{figure}[!ht]
\centering \psfull
 \epsfig{file=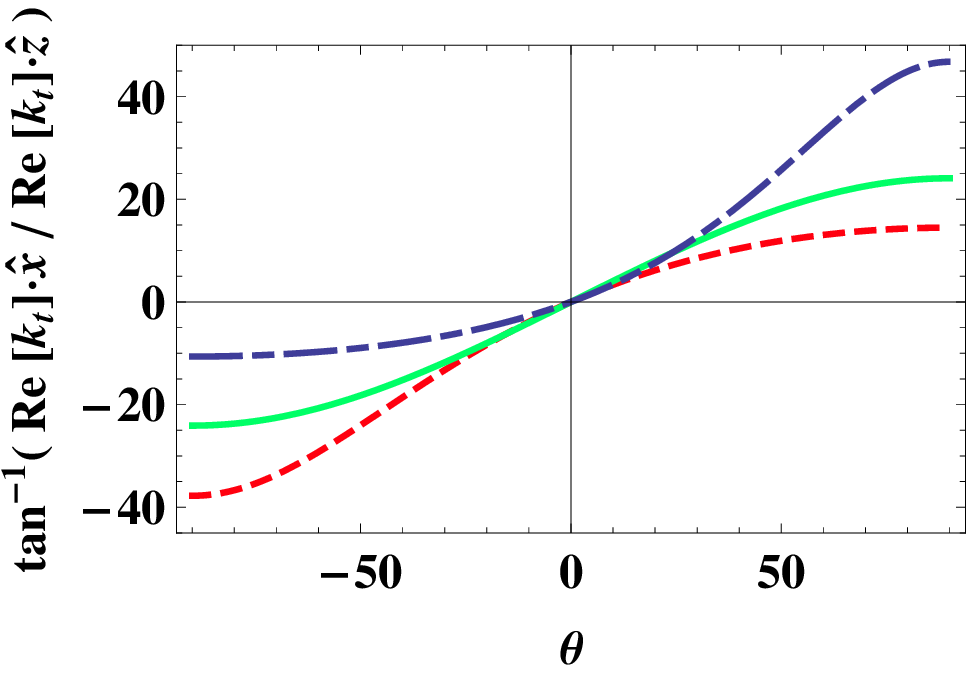,width=3.1in} \hfill
\epsfig{file=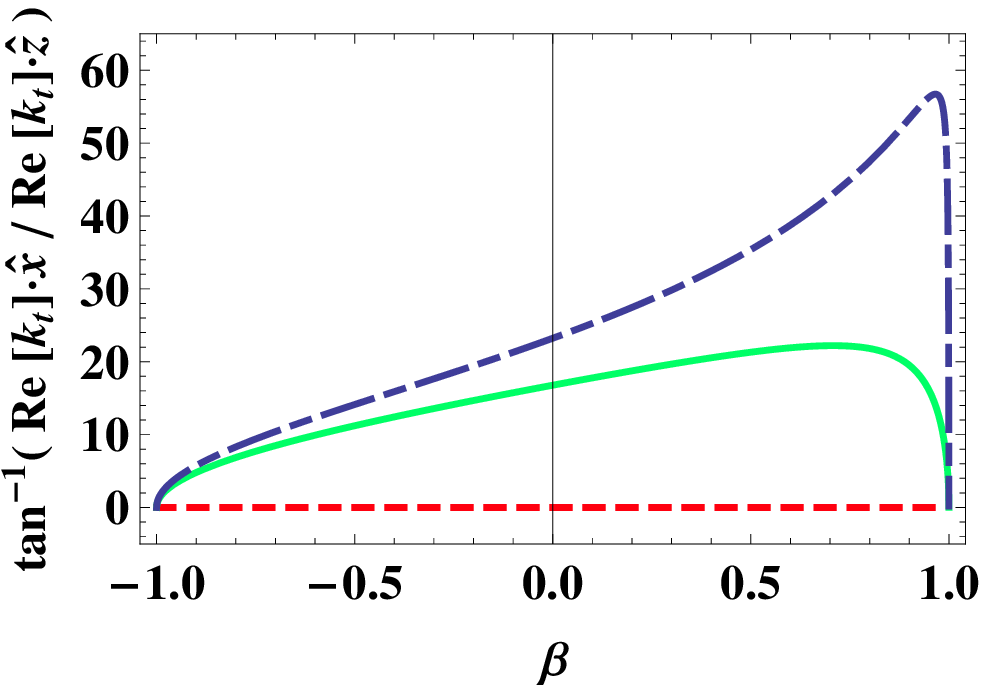,width=3.1in}
 \caption{\label{kv_e1} The orientation angle of
 $\mbox{Re} \, \#k_{t}$ (in degree)  plotted against angle of incidence $\theta \in \le -90^\circ, 90^\circ
 \ri$
 for  $\beta = -0.5$   (broken curve,  red), $\beta = 0$
(solid curve,  green), and $\beta = 0.7 $ (broken dashed curve,
blue); and plotted against relative speed $\beta \in \le -1, 1
 \ri$
 for  $\theta = 0^\circ$   (broken curve,  red), $\theta =  45^\circ$
(solid curve,  green), and $\theta =  75^\circ $ (broken dashed
curve, blue).
 The refracting half--space is occupied by material A ($\eps' =  6 + 0.05 i $).}
\end{figure}

For the purposes of illustration, let us suppose firstly that the
material occupying the half--space $z>0$ has a relative permittivity
$\eps' =  6 + 0.05 i $. We call this material A. We note that the
case for the corresponding nondissipative material, specified by a
real--valued relative permittivity of 6, was investigated previously
using an approach based on the Minkowski constitutive relations
\c{Hiding}. The orientation angle of the real part of
$\#k_t$, as defined by $\tan^{-1} \le \kappa
/ \mbox{Re} \, k_{zt} \ri$,
 is plotted in Fig.~\ref{kv_e1} versus (a)  angle of incidence
  $\theta$ for the relative speeds $\beta \in \lec -0.5, \, 0, \, 0.7
\ric$; and (b)  relative speed $\beta$ for the  angles of incidence
$\theta \in \lec 0^\circ, \, 45^\circ, \, 75^\circ \ric$. We see
that: (a) for a fixed angle of incidence $\theta$, the orientation
angle of $\mbox{Re} \, \#k_t$ increases as $\beta$ increases (with
the exception of the case of normal incidence wherein the
orientation angle of $\mbox{Re} \, \#k_t$ is $0^\circ$ for all
values of $\beta$); and (b)
 for a fixed relative speed $\beta$, the orientation angle of
$\mbox{Re} \, \#k_t$ increases as $\theta$ increases. Significantly,
$\tan^{-1} \le \kappa / \mbox{Re} \, k_{zt} \ri > 0 \; (< 0)$ for
all values of $\beta$, provided that $\theta > 0 \; (< 0)$. Thus, we
see that  refraction at the interface  $z=0$ is always
positive, regardless of the values of $\beta$ and $\theta$.

\begin{figure}[!ht]
\centering \psfull \epsfig{file=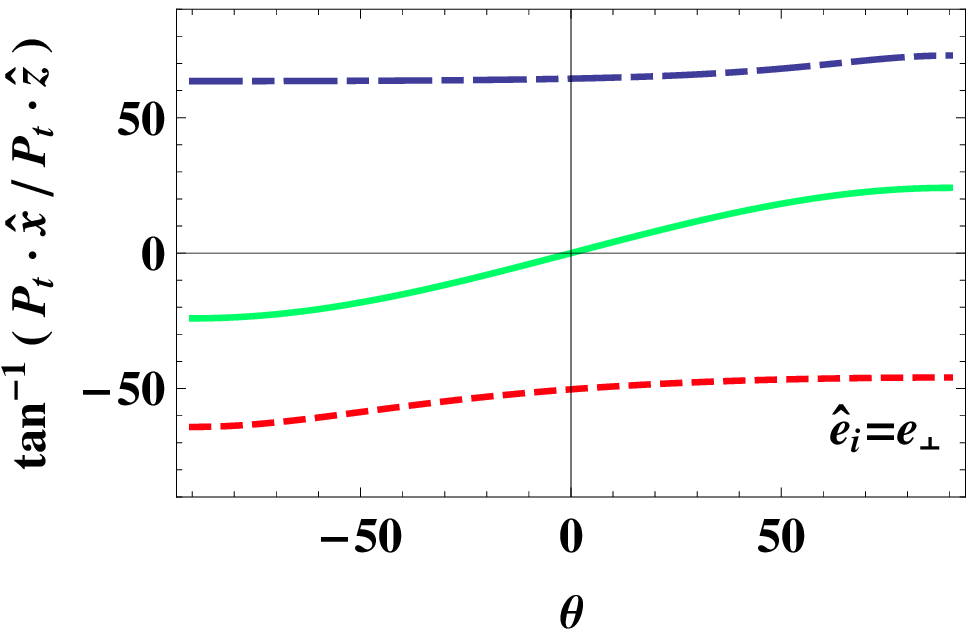,width=3.1in} \hfill
 \epsfig{file=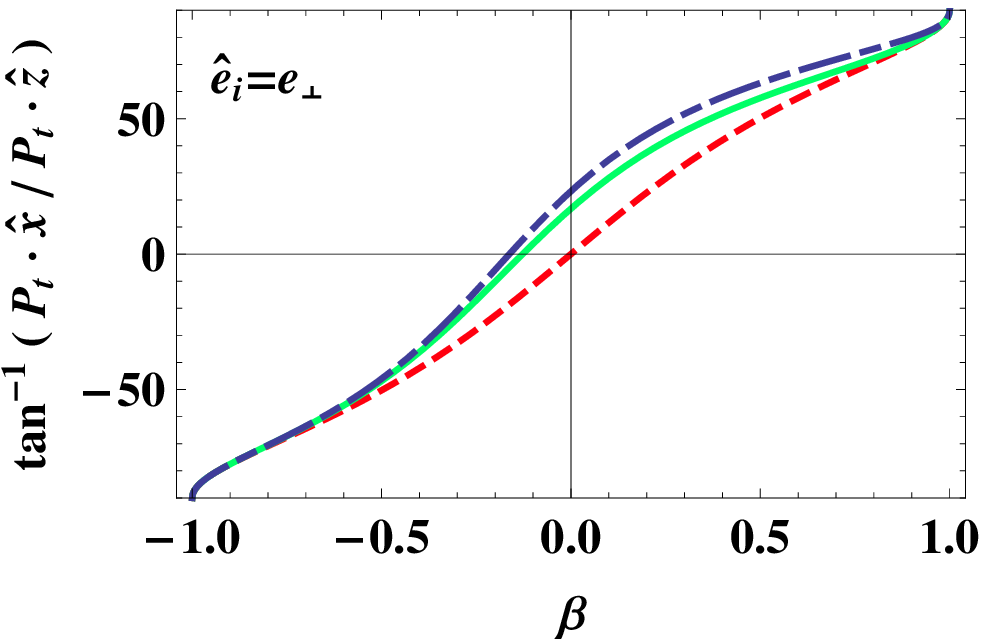,width=3.1in}\\
\epsfig{file=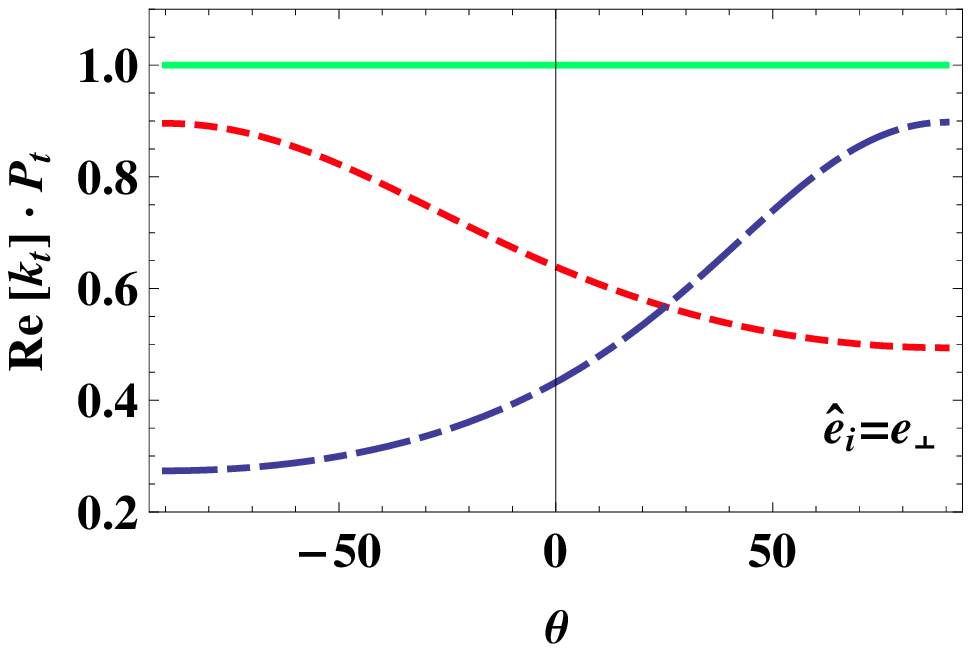,width=3.1in} \hfill
\epsfig{file=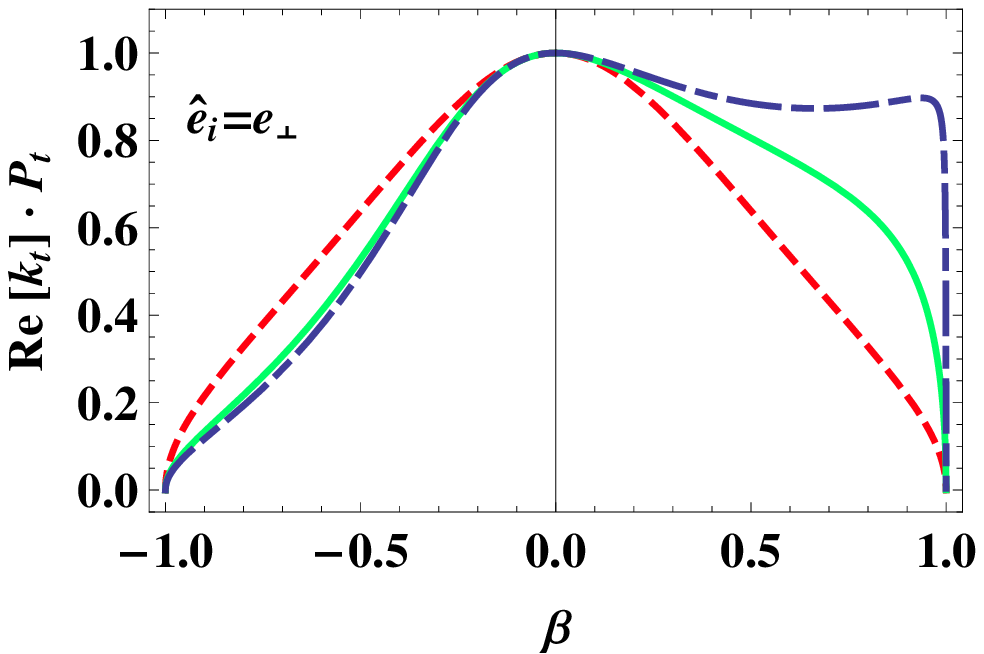,width=3.1in}
 \caption{\label{pv_e1} The orientation angle of
 the cycle--averaged Poynting vector $\#P_{\,t}$ and the quantity $ \#P_{t} \. \mbox{Re}
\, \#k_{t} $ (normalized) plotted against angle of incidence $\theta
\in \le -90^\circ, 90^\circ \ri$
 for  $\beta = -0.5$   (broken curve,  red), $\beta = 0$
(solid curve,  green), and $\beta = 0.7 $ (broken dashed curve,
blue); and plotted against relative speed $\beta \in \le -1, 1
 \ri$
 for  $\theta = 0^\circ$   (broken curve,  red), $\theta =  45^\circ$
(solid curve,  green), and $\theta =  75^\circ $ (broken dashed
curve, blue).
 The refracting half--space is occupied by material A ($\eps' =  6 + 0.05 i $).
 Plots are shown for the incident $s$-polarization state.}
\end{figure}

Next we turn to energy flux in the  half--space $z>0$, as provided by
the cycle--averaged Poynting vector $\#P_t$. In the limit $| \#r |
\to 0$,  we have \c{Chen}
\begin{eqnarray}
 \left. \#P_{t} \, \right|_{\#r= \#0}  &=&    \frac{1}{2} \les \le \mbox{Re} \,
 \#e_{t} \ri
\times \le \mbox{Re} \, \#h_{t} \ri  + \le \mbox{Im} \, \#e_{t} \ri
\times \le \mbox{Im} \, \#h_{t} \ri \ris.
\end{eqnarray}
The case of incident $s$--polarization state for which
\begin{equation}
\#e_i = a_s \, \#e_\perp \equiv a_s \, \hat{\#y} \l{s_polar}
\end{equation}
is distinguished from the  case of incident $p$--polarization state
described by
\begin{equation}
\#e_i = a_p \,  \#e_\parallel \equiv a_p \le -   \cos \theta \,
\hat{\#x} + \sin \theta \, \hat{\#z} \ri ,
\end{equation}
with $a_{s,p}$ being complex--valued amplitudes. In
Fig.~\ref{pv_e1}, the orientation  angle of the cycle--averaged
Poynting vector $\#P_t$ evaluated in the limit $| \#r | \to 0$,  as
defined by $ \left. \tan^{-1} \le \#P_t \. \hat{\#x} / \#P_t \.
\hat{\#z} \ri \right|_{\,\#r = \#0}$,
 is plotted  versus (a) the
angle of incidence $\theta$ for the relative speeds $\beta \in \lec
-0.5, \, 0, \, 0.7 \ric$, and (b)  the relative speed $\beta$ for the
 angles of incidence $\theta \in \lec 0^\circ, \, 45^\circ, \, 75^\circ \ric$.
Also presented in this figure are the corresponding plots of the
quantity $\left. \#P_{t} \, \right|_{\#r= \#0} \.  \mbox{Re} \,
\#k_t $ which determines whether the phase velocity is positive or
negative. The graphs in Fig.~\ref{pv_e1} are for the incident plane wave having
the $s$--polarization state; the corresponding graphs for the $p$--polarized incident plane wave
are almost (but not exactly) identical. The
orientation angle of $ \left. \#P_{t} \, \right|_{\#r= \#0} $ is
observed to (a) increase as  $\beta$ increases, at a fixed angle of
incidence; and (b) increase as
 $\theta$ increases, at a fixed relative speed.
Furthermore, by comparing Figs.~\ref{kv_e1} and \ref{pv_e1}, it is
apparent that  counterposition of $ \left. \#P_{t} \, \right|_{\#r=
\#0}$ and $ \mbox{Re} \, \#k_t $ occurs when $ - \beta$ is
sufficiently large for $\theta > 0$, and when $  \beta$ is
sufficiently large for $\theta < 0$. Counterposition does not occur
at all when $\beta = 0$, regardless of the angle of incidence. In
addition, the phase velocity of the refracted plane wave is positive
for all values of $\beta$ and $\theta$.

Let us now consider a second numerical example where the dissipative
material occupying $z>0$ possesses the relative permittivity $\eps'
= -0.34  + 0.04 i $. This material, which we call material B,  has
the relative permittivity of aluminium at the free--space wavelength
$\lambdao = 103$ nm \c{Ditchburn}. Material B is radically different
from material A insofar as the former can support NPV propagation at rest, as
we demonstrate in the following.

\begin{figure}[!ht]
\centering \psfull
 \epsfig{file=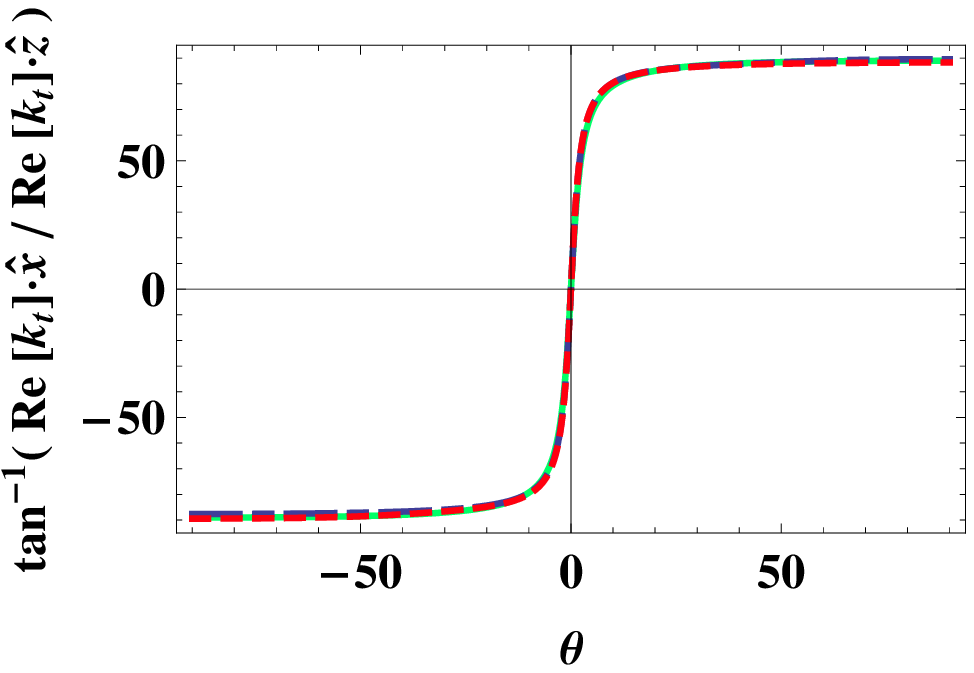,width=3.1in} \hfill
 \epsfig{file=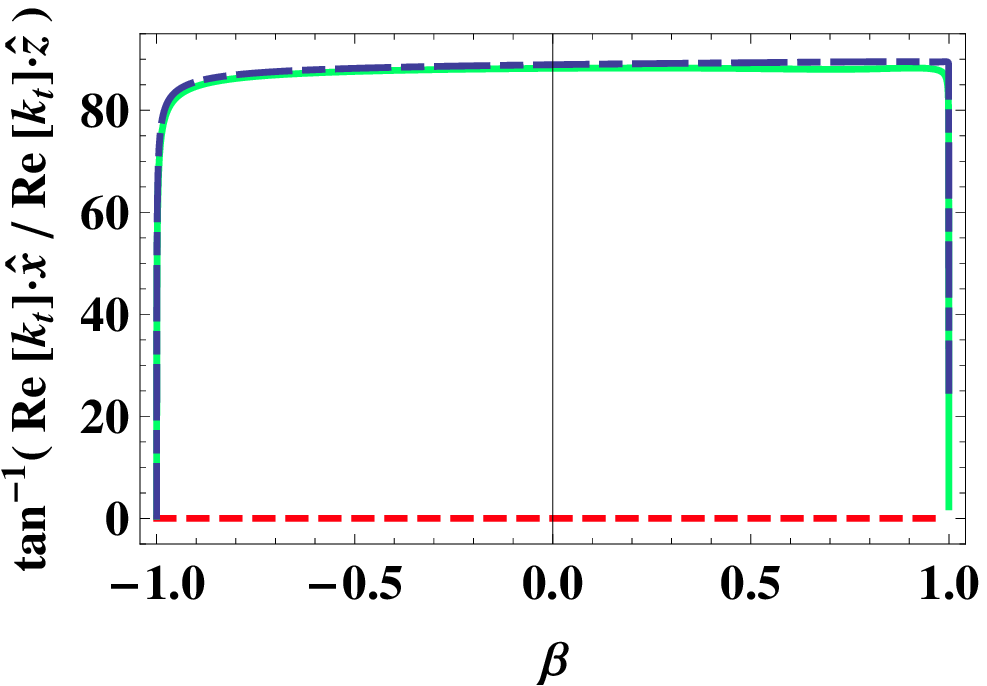,width=3.1in}
 \caption{\label{kv_e2}
As Fig.~\ref{kv_e1} except that material A is replaced by material B
 ($\eps' = - 0.34 + 0.04 i $).}
\end{figure}

The orientation angle of the real part of  $\#k_t$
 is plotted in Fig.~\ref{kv_e2} versus (a) the angle of incidence
  $\theta$ for the relative speeds $\beta \in \lec -0.5, \, 0, \, 0.7
\ric$, and (b) the  relative speed $\beta$ for the angles of incidence
$\theta \in \lec 0^\circ, \, 45^\circ, \, 75^\circ \ric$. We see
that: (a)  the orientation angle of $\mbox{Re} \, \#k_t$ increases
uniformly as $\theta$ increases, at approximately the same rate
regardless of the value of $\beta$; and (b)
 for $90^\circ > \theta \gtrapprox 10^\circ$,
the orientation angle of $\mbox{Re} \, \#k_t$ is nearly $90^\circ$,
and for $- 90^\circ < \theta \lessapprox - 10^\circ$, the
orientation angle of $\mbox{Re} \, \#k_t$ is nearly $- 90^\circ$,
regardless of the value of $\beta$. And, in the limits $\beta \to
\pm 1$, the  orientation angle of $\mbox{Re} \, \#k_t$ tends to
$0^\circ$. Notably,  the  orientation angle of $\mbox{Re} \, \#k_t$
is always $0^\circ$ for the case of normal incidence, at all values
of $\beta \in \le -1, 1 \ri$. As in the previous example,
refraction at the interface $z=0$ is positive
for all values of $\beta$ and $\theta$.

\begin{figure}[!ht]
\centering \psfull \epsfig{file=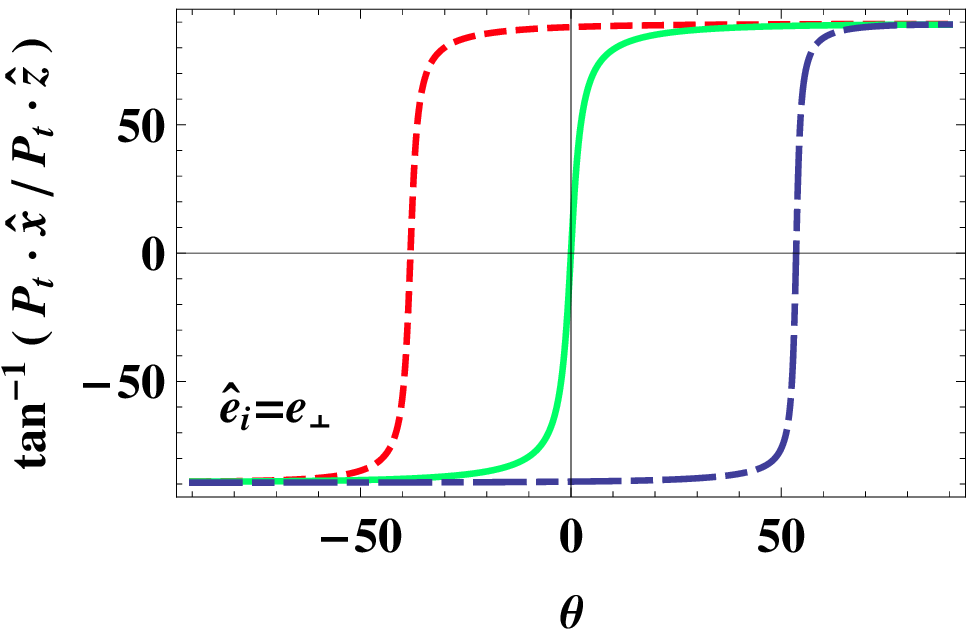,width=3.1in} \hfill
 \epsfig{file=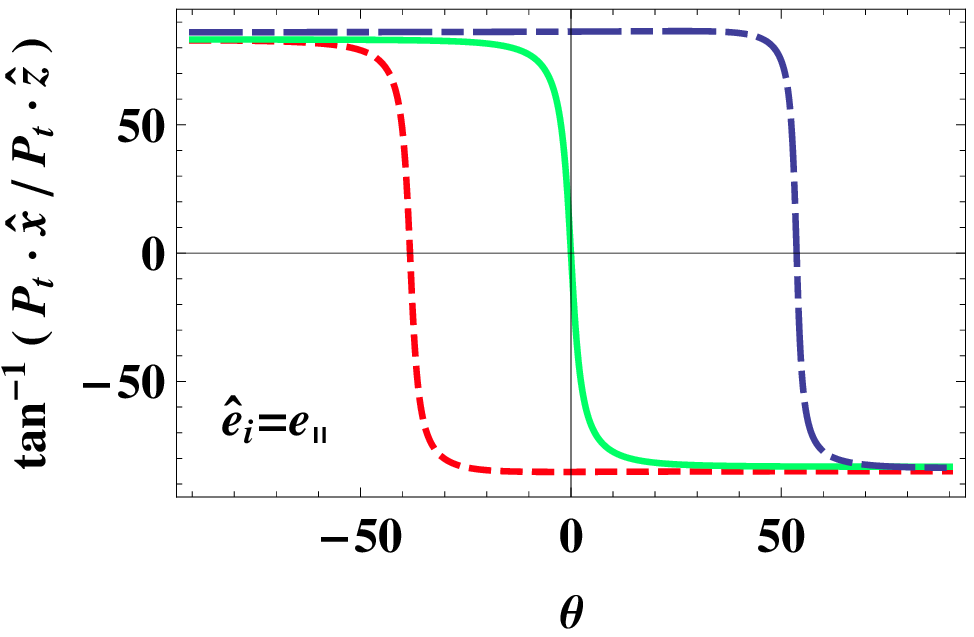,width=3.1in}\\
 \epsfig{file=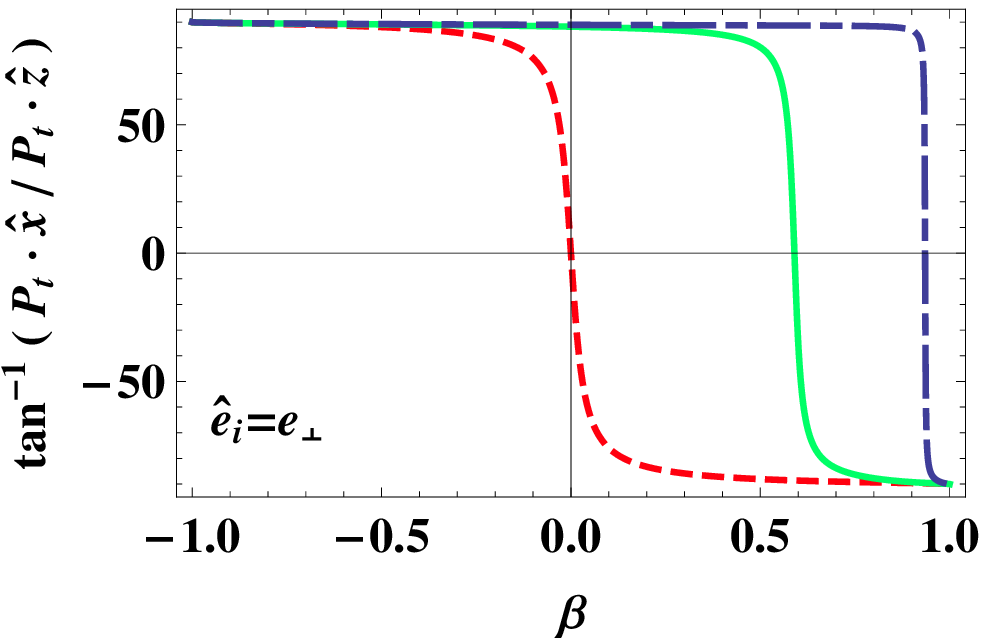,width=3.1in} \hfill
 \epsfig{file=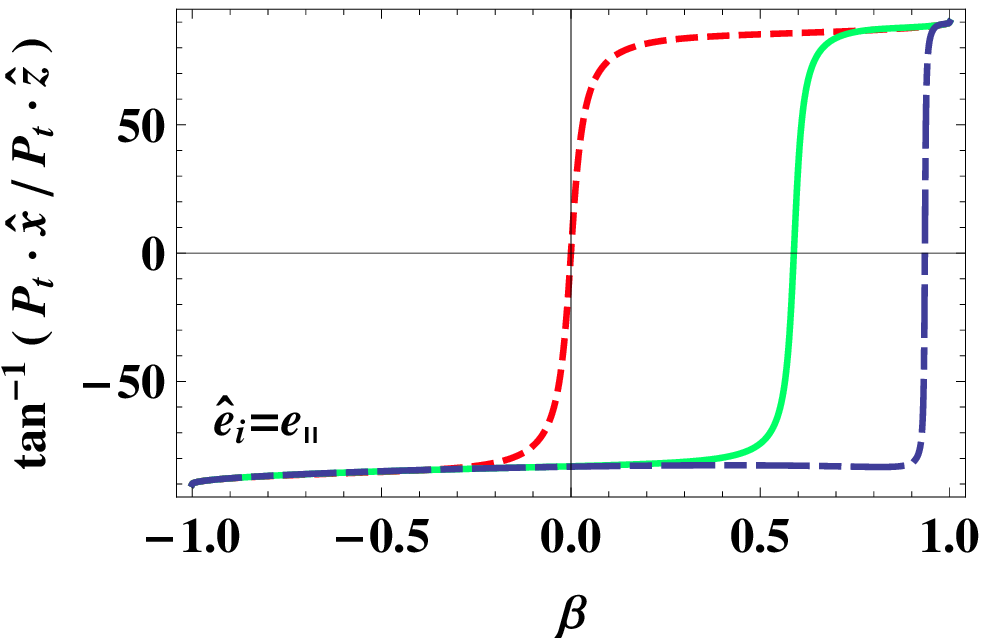,width=3.1in}\\
\epsfig{file=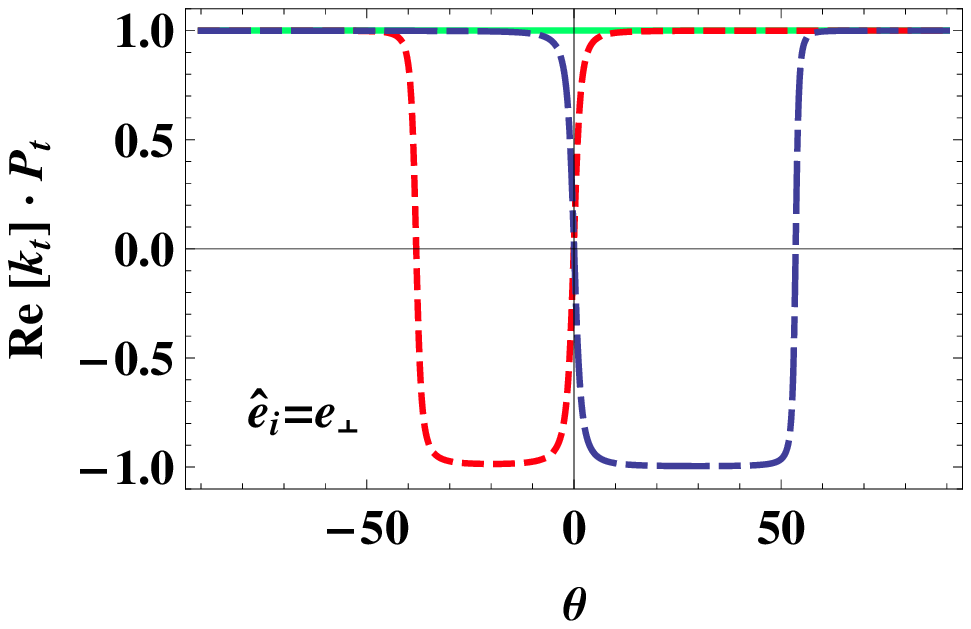,width=3.1in} \hfill
\epsfig{file=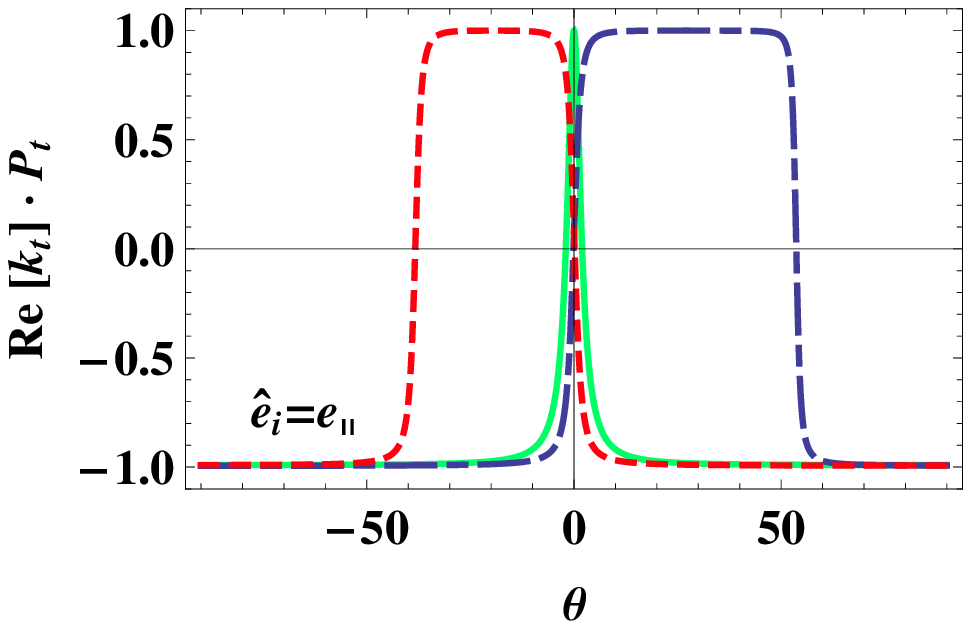,width=3.1in}
\\
\epsfig{file=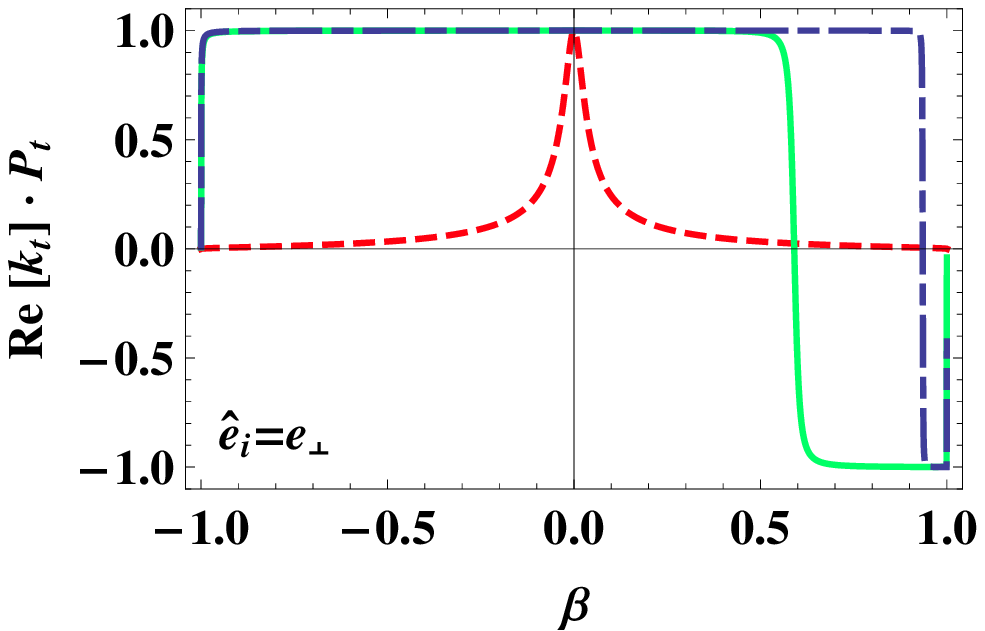,width=3.1in} \hfill
\epsfig{file=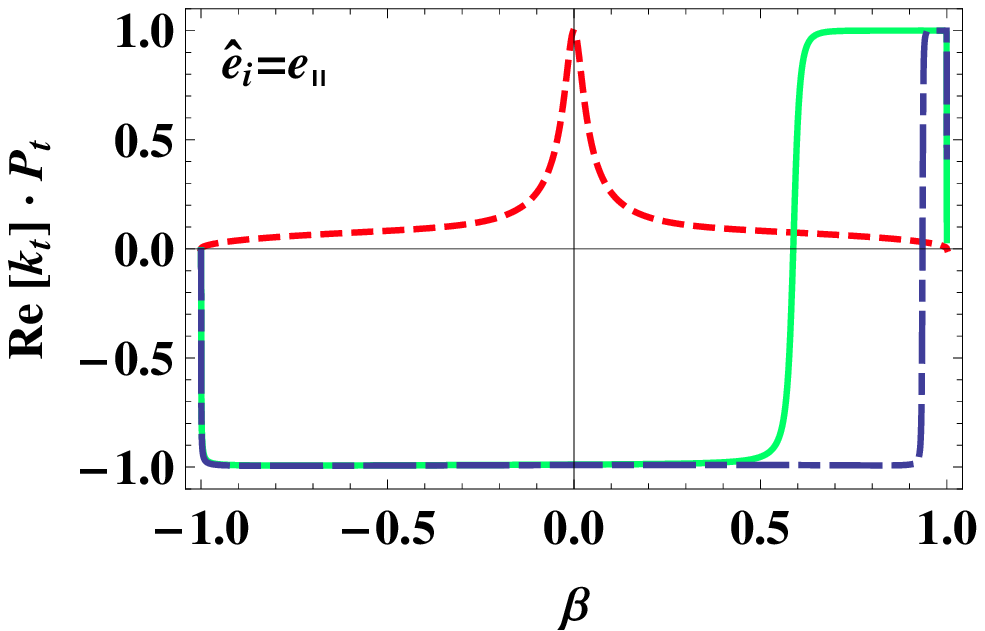,width=3.1in}
 \caption{\label{pv_e2}
As Fig.~\ref{pv_e1} except that material A is replaced by material B
 ($\eps' = - 0.34 + 0.04 i $), and here the plots for the incident  $s$-polarization state (left)
 are quite different to the corresponding plots for the incident $p$-polarization  state (right).
}
\end{figure}

Graphs of the orientation angle of the cycle--averaged Poynting
vector evaluated in the limit $| \#r | \to 0$, analogous to those in
Fig.~\ref{pv_e1}, are presented in Fig.~\ref{pv_e2}. Unlike the case
for material A, graphs for the incident $s$- and $p$-polarization
states are quite different here. The  orientation angle of $ \left.
\#P_{t} \, \right|_{\#r= \#0} $ is observed to be  nearly $\pm
90^\circ$ for most values of $\beta \in \le -1, 1\ri$ and for
$\theta \in \le - 90^\circ, 90^\circ \ri$. There exist: (a) a
narrow range of values of $\theta$ for a fixed value of $\beta$
(that is centered on $\theta = 0^\circ$ for $\beta = 0$); and (b)
  a narrow range of values of $\beta$ (that is
centered on $\beta = 0$ for $\theta = 0^\circ$) for which the
orientation angle of $ \left. \#P_{t} \, \right|_{\#r= \#0} $ passes
through $ 0^\circ$, as it increases (or decreases) from nearly $ -
90^\circ$ (or $ 90^\circ$) to nearly $  90^\circ$ (or $ -
90^\circ$). Counterposition of $ \left. \#P_{t} \, \right|_{\#r=
\#0}$ and $ \mbox{Re} \, \#k_t $ occurs at $\beta = 0$ for the
incident  $p$-polarization state  but not for the incident  $s$-polarization state, for
all angles of incidence except $\theta = 0^\circ$ (when  $ \left.
\#P_{t} \, \right|_{\#r= \#0}$ and $ \mbox{Re} \, \#k_t $ are both
directed normally away from the interface). For the incident $s$-polarization
state, counterposition occurs when $\beta > 0$ for $\theta =
0^\circ$, $\beta > 0.6$ for $\theta = 45^\circ$, and $\beta > 0.92$
for $\theta = 75^\circ$. For the incident $p$-polarization state,
counterposition occurs when $\beta < 0$ for $\theta = 0^\circ$,
$\beta < 0.6$ for $\theta = 45^\circ$, and $\beta < 0.92$ for
$\theta = 75^\circ$.

The sign of the phase velocity can be inferred from the
corresponding plots $\left. \#P_{t} \, \right|_{\#r= \#0} \.
\mbox{Re} \, \#k_t $ which are also presented in
 Fig.~\ref{pv_e2}.
The phase velocity of the refracted plane wave is positive at $\beta
= 0$  for all angles of incidence, when the incident plane wave is
$s$ polarized. However,  when the incident plane wave is $p$
polarized,  the refracted plane wave has NPV at $\beta = 0$  for
$\theta < -3^\circ$ and $\theta
> 3^\circ$.
 For the incident $s$-polarization
state, NPV arises in the refracted field when $\beta > 0.6$ for
$\theta = 45^\circ$, and $\beta > 0.92$ for $\theta = 75^\circ$. For
the incident $p$-polarization state, NPV arises in the refracted
field when $\beta < 0.6$ for $\theta = 45^\circ$, and $\beta < 0.92$
for $\theta = 75^\circ$.

\section{Beam propagation through  a moving slab} \l{moving_slab}

\subsection{Theoretical considerations}

Suppose now that  the uniformly moving material which occupied the
half--space $z>0$
  in \S\ref{moving_hs}  is
  restricted in its extent  to a slab region
occupying  $0 < z < L$. As before, the uniformly moving  material
 is described by the constitutive relations
\r{Con_rel_p} in the inertial reference frame $\Sigma'$. The regions
$z > L$ and $z < 0 $ are vacuous.

With respect to the  inertial reference frame $\Sigma$, a
two--dimensional beam with electric field phasor \c{Haus}
\begin{equation}
\#E_i   = \int^{\infty}_{-\infty}
  \#e_i (\psi) \, \Psi (\psi)   \, \exp \les i
  \le \#k_i \. \#r  - \omega t \ri
 \ris \; d \psi , \hspace{25mm} z \leq 0\l{Ei2}
\end{equation}
is incident upon the uniformly
moving slab at a mean angle $\theta$
relative to the slab's thickness direction $\mbox{\boldmath$\hat{z}$}$.
The beam is an angular spectrum of plane waves, with
 each planewave contributor having the wavevector
\begin{equation}
\#k_i = \ko \les \le \psi \, \cos \theta + \sqrt{1-\psi^2} \, \sin
\theta \ri \hat{\#x} - \le \psi \, \sin \theta - \sqrt{1-\psi^2} \,
\cos \theta \ri \hat{\#z} \, \ris.
\end{equation}
A Gaussian form
\begin{equation}
\Psi (\psi) = \frac{\ko \, \wo}{\sqrt{2 \pi}} \, \exp \les
-\frac{\le \ko \, \wo \, \psi \ri^2}{2}\ris
\end{equation}
is adopted for the angular--spectral function $\Psi (\psi)$, with
$\wo$ being the width of the beam waist \c{Haus}. Two polarization
states are considered: the $s$-polarization state described by \r{s_polar}
and the $p$-polarization state for which
\begin{equation}
\#e_i (\psi) = a_p \, \#e_\parallel  \equiv a_p \les  \le \psi \,
\sin \theta - \sqrt{1-\psi^2} \, \cos \theta \ri \hat{\#x} + \le
\psi \, \cos \theta + \sqrt{1-\psi^2} \, \sin \theta \ri \hat{\#z}
\ris.
\end{equation}

As the velocity is tangential to the interfaces and lies wholly in
the plane of incidence, the reflected and transmitted beams  have
spatial Fourier representations similar to that of the incident beam
given by \r{Ei2}.  In  $\Sigma$,
 the reflected  beam
is represented by the electric field  phasor
\begin{equation}
\#E_r   = \int^{\infty}_{-\infty}
  \#e_r (\psi) \, \Psi (\psi)   \, \exp \les i
  \le \#k_r \. \#r  - \omega t \ri
 \ris \; d \psi, \hspace{25mm} z \leq 0, \l{Er}
\end{equation}
with the wavevector of each planewave contributor being
\begin{equation} \#k_r = \ko \les \le \psi \, \cos
\theta + \sqrt{1-\psi^2} \, \sin \theta \ri \hat{\#x} + \le \psi \,
\sin \theta - \sqrt{1-\psi^2} \, \cos \theta \ri \hat{\#z} \, \ris
\end{equation}
and the corresponding amplitudes
\begin{equation}
\#e_r (\psi) = \left\{
\begin{array}{ccr}
r_s \,\#e_\perp & \mbox{for} &  \#e_i (\psi) = \#e_\perp,
\vspace{6pt}
\\
r_p \Big[ - \le \psi \, \sin \theta - \sqrt{1-\psi^2} \, \cos \theta
\ri \hat{\#x}&& \\ + \le \psi \, \cos \theta + \sqrt{1-\psi^2} \,
\sin \theta \ri \hat{\#z} \Big] & \mbox{for} & \#e_i (\psi) =
\#e_\parallel. \end{array} \right.
\end{equation}
Likewise, in $\Sigma$, the
transmitted  beam is represented by the electric field  phasor
\begin{equation}
\#E_\tau   = \int^{\infty}_{-\infty}
  \#e_\tau (\psi) \, \Psi (\psi)   \, \exp \lec i
  \les \#k_\tau \. \le \#r - L \hat{\#z} \ri  - \omega t \,  \ris
 \ric \; d \psi, \hspace{25mm} z \geq L, \l{Et}
\end{equation}
with $\#k_\tau = \#k_i$ and the amplitudes
\begin{equation}
 \#e_\tau (\psi) = \left\{
\begin{array}{ccr}
\tau_s \, \#e_\perp  & \mbox{for} &  \#e_i (\psi) = \#e_\perp ,
\vspace{6pt} \\
\tau_p \, \#e_i (\psi) & \mbox{for} & \#e_i (\psi) = \#e_\parallel.
\end{array}
\right.
\end{equation}

The electric field phasors for the reflected and transmitted beams
in   $\Sigma$ are calculated via the strategy
outlined in \S\ref{moving_hs}. That is, first the problem is
transformed to the inertial reference frame  $\Sigma'$ wherein the
boundary-value problem is solved for each planewave contributor to
the incident beam, using expressions for the reflection coefficients $r_{s,
p}$ and transmission coefficients $\tau_{s, p}$ appropriate for
$\Sigma'$
\c{Hiding}. Then the electric phasors for the contributor plane
waves are transformed back to   $\Sigma$, and
integrated with respect to $\psi$ to yield $\#E_{r,\tau}$.

\subsection{Numerical studies}

\begin{figure}[!ht]
\centering \psfull \epsfig{file=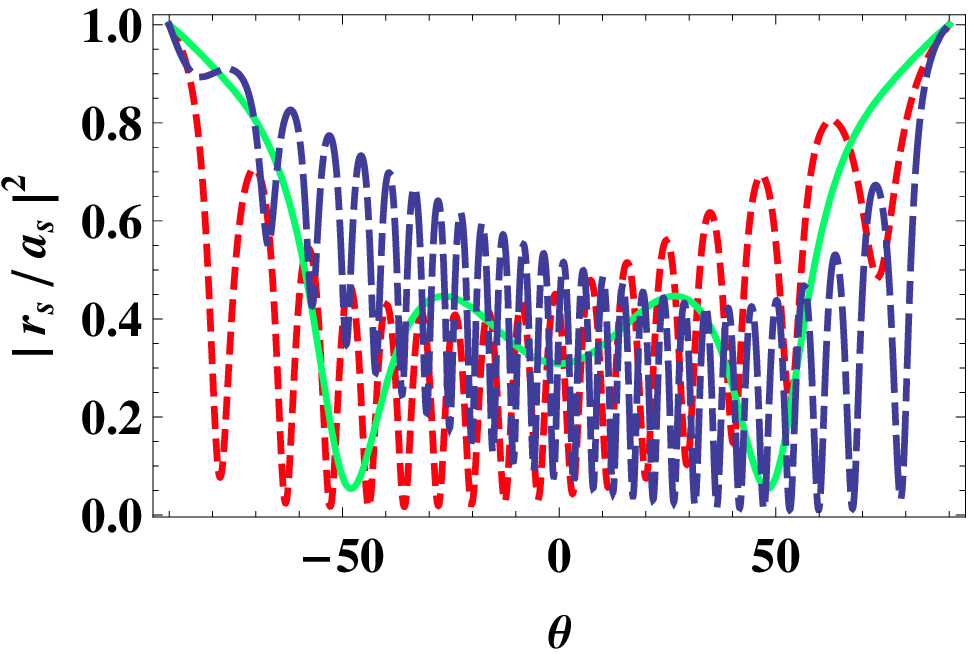,width=3.1in} \hfill
\epsfig{file=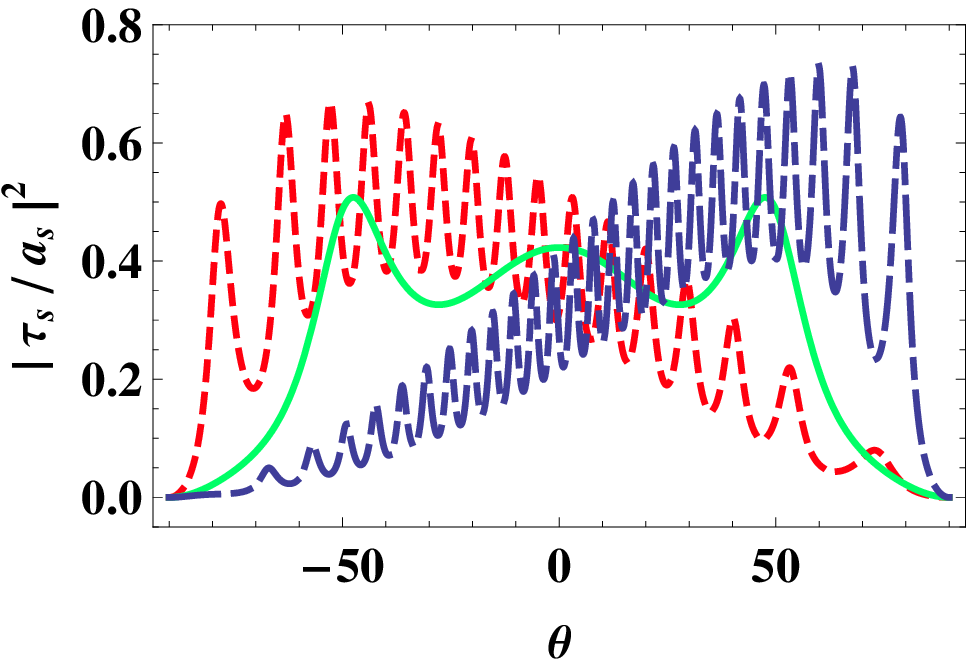,width=3.1in}
 \caption{\label{trans_e1} The
reflectance $\left| r_{s} / a_{s} \right|^2$ and transmittance
$\left| \tau_{s} / a_{s} \right|^2$,   plotted against angle of
incidence $\theta \in \le -90^\circ, 90^\circ \ri$,  for $\beta =
-0.5$   (broken curve,  red), $\beta = 0$ (solid curve, green), and
$\beta = 0.7 $ (broken dashed curve, blue). The
 moving slab of thickness $L  = 3 \lambdao$ is made of material A ($\eps' =  6 + 0.05 i $).
 }
\end{figure}

The slab thickness was fixed at $L = 3 \lambdao$. Following
\S\ref{num_1}, we begin by considering a slab made of material A
($\eps' = 6 + 0.05i$). In Fig.~\ref{trans_e1}, the reflectance
$\left| r_{s} / a_{s} \right|^2$ and transmittance $\left| \tau_{s}
/ a_{s} \right|^2$,   are plotted against the angle of incidence
$\theta \in \le -90^\circ, 90^\circ \ri$, for the relative speeds
$\beta \in \lec -0.5, \, 0, \, 0.7 \ric$. The reflectance at the
angles of incidence $\theta $ and $- \theta $ are the same for
$\beta = 0$, but this is not true for $\beta \neq 0$; and the same
is true for the  transmittance. The reflectances
 and   transmittances oscillate markedly as $\theta$ increases.
 This effect~---~which is due to multiple reflections at the $z=0$ and
$z=L$ interfaces~---~disappears as the slab thickness increases (as
we confirmed in calculations not displayed here). The graphs for
 the reflectance
$\left| r_{p} / a_{p} \right|^2$ and transmittance $\left| \tau_{p}
/ a_{p} \right|^2$ (which are not displayed here) are similar to
those for $\left| r_{s} / a_{s} \right|^2$ and $\left| \tau_{s} /
a_{s} \right|^2$.

For a mean angle of incidence $\theta = 45^\circ$, with the incident beam focussed on
$\left\{x=0,z=0\right\}$ with  waist $\wo = \lambdao$, we investigated
  the reflected and  transmitted beams at the relative speeds
   $\beta \in \lec - 0.5,\,  0, \,
   0.7 \ric$.
The energy density
\begin{equation}
 |\#E |^2 = \left\{
 \begin{array}{lcr}
|\#E_i   + \#E_r |^2 & \mbox{for} & z \leq 0, \vspace{4pt} \\
|\#E_\tau |^2 & \mbox{for} & z \geq L.
\end{array}
\right.
\end{equation}
was used for this purpose.

\begin{figure}[!ht]
\centering \psfull \epsfig{file=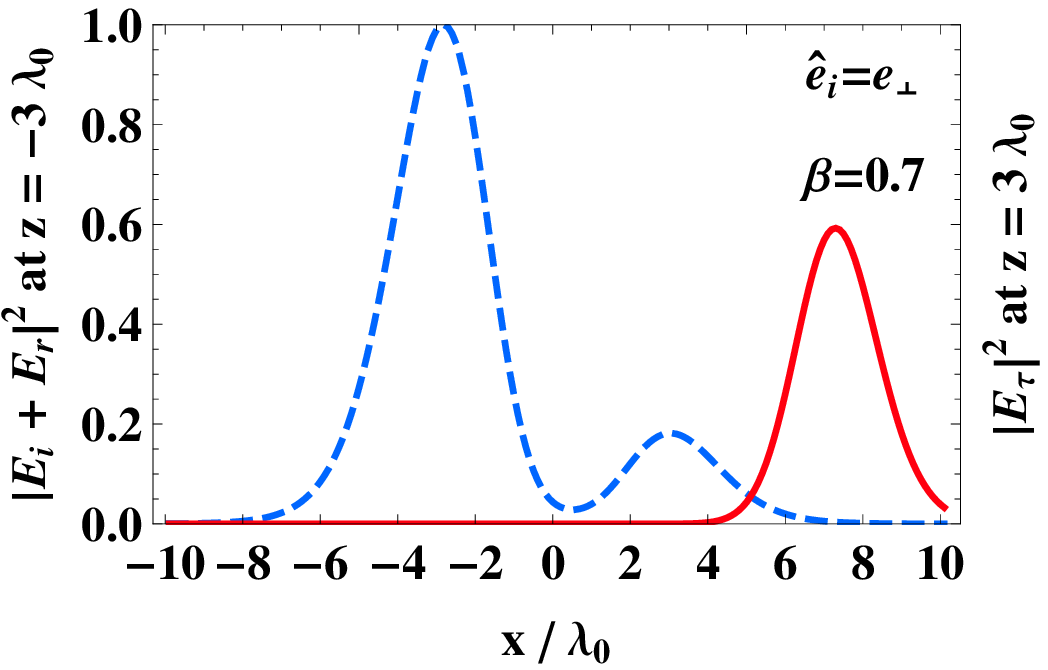,width=3.1in}
\hfill
\epsfig{file=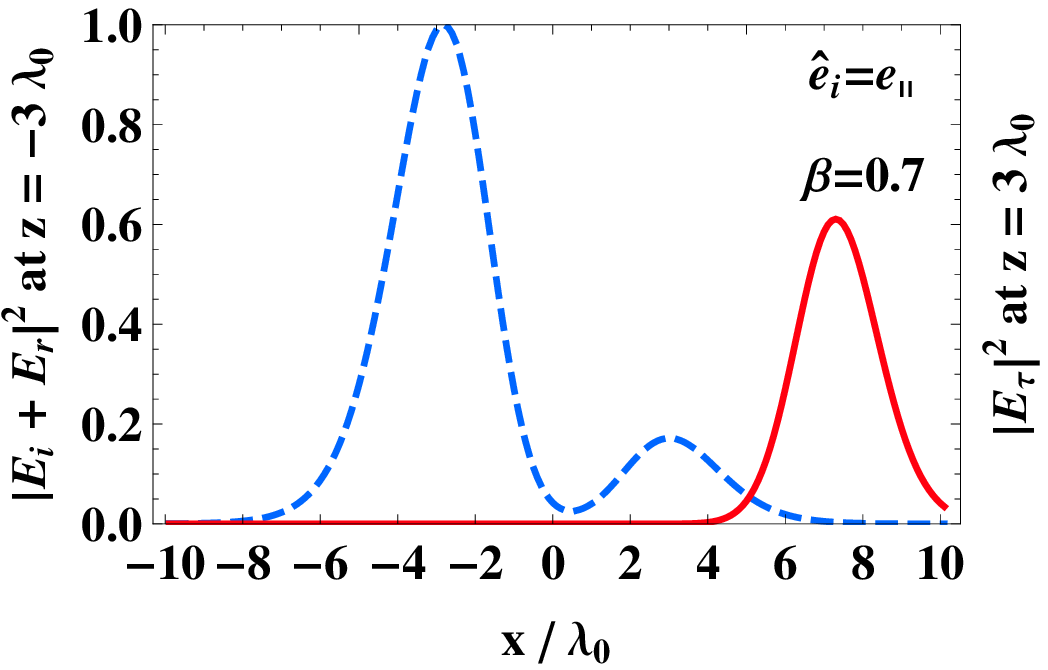,width=3.1in} \\
\epsfig{file=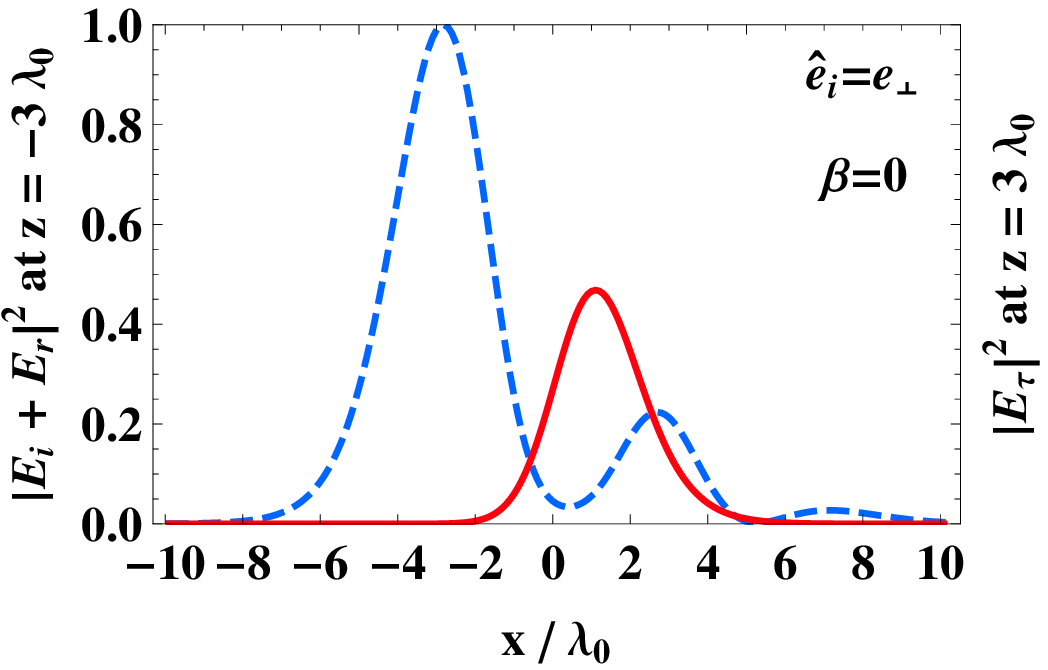,width=3.1in}  \hfill
 \epsfig{file=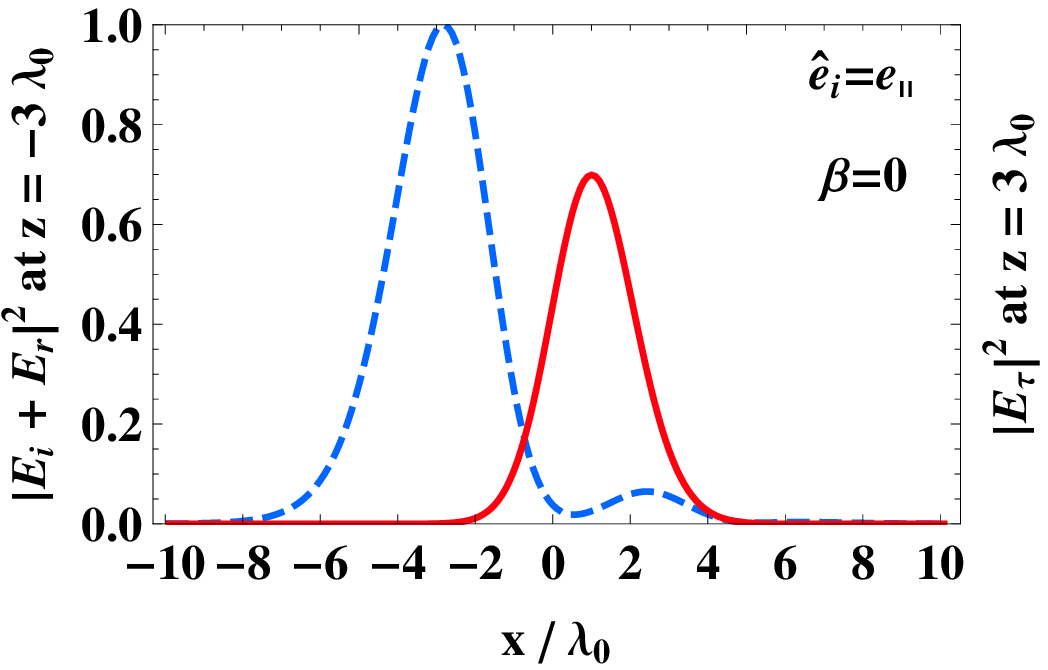,width=3.1in}
  \\
\epsfig{file=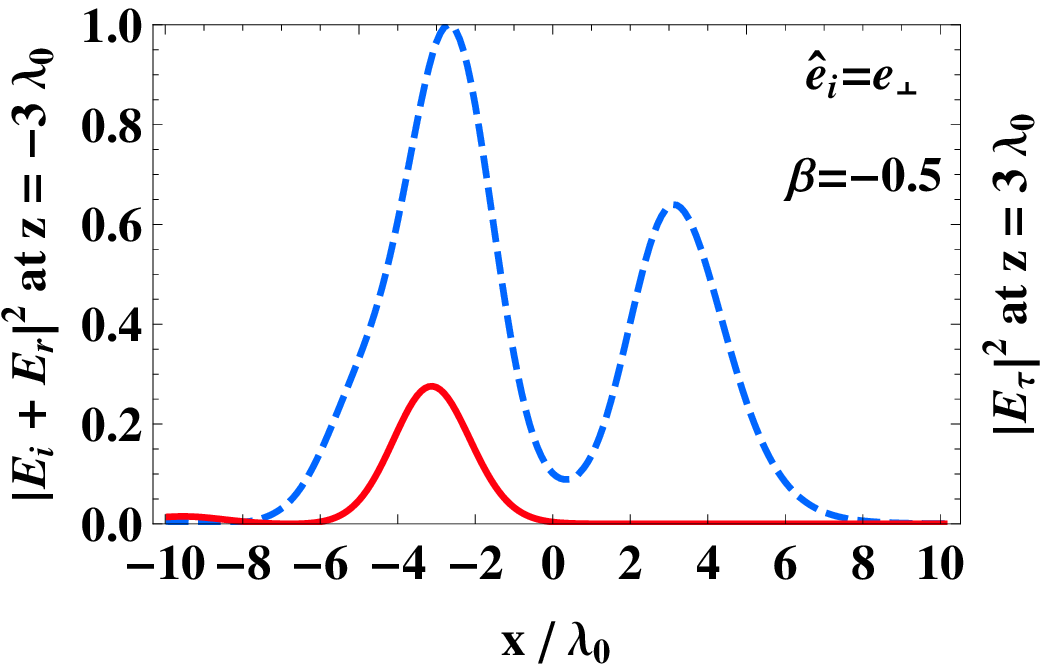,width=3.1in}  \hfill
 \epsfig{file=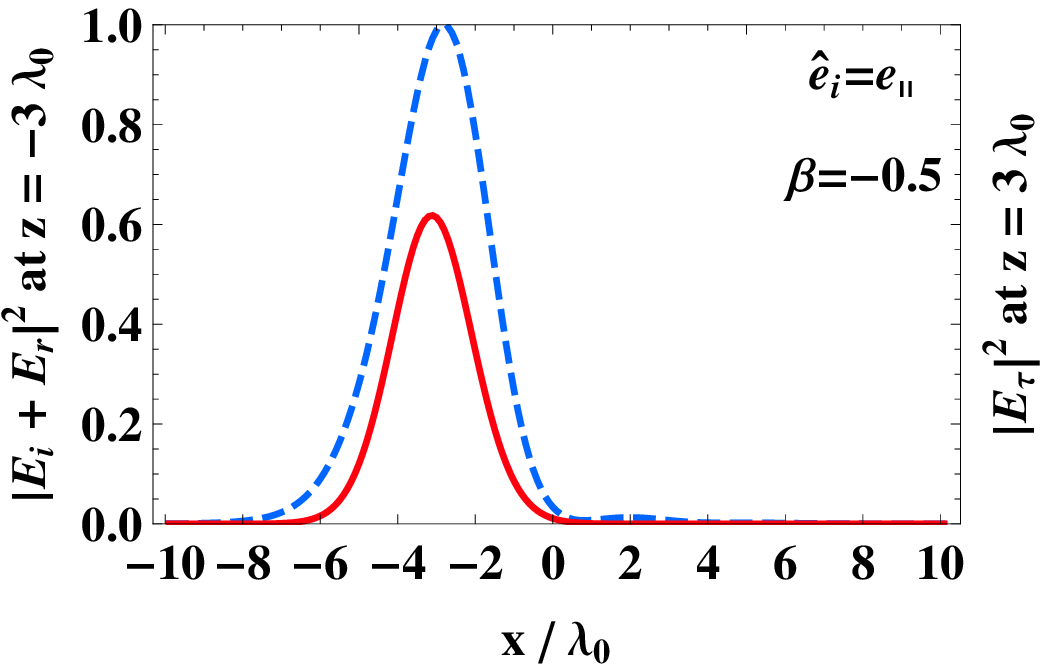,width=3.1in}
 \caption{\label{energy_e1} The energy densities at $z =  3
 \lambdao$ (solid curves,  red) and  $z = - 3
  \lambdao$ (broken curves,  blue)
 plotted against $x \in \le -10 \lambdao, 10 \lambdao \ri$
for $\beta = 0.7$ (top), $\beta = 0$ (middle), and $\beta = -0.5$
(bottom), for an incident $s$-polarized beam (left) and an incident
$p$-polarized beam (right). The
 moving slab of thickness $L  = 3 \lambdao$ is made of material A ($\eps' =  6 + 0.05 i $).
The centre of the incident beam impinges on the moving slab at the
point $x=0$, $z=0$, at an angle of incidence  $\theta = 45^\circ$.
   }
\end{figure}

 This energy densities at the planes  $z= -L$ and $z=L$ are plotted versus $ x
\in \le -10 \lambdao , 10 \lambdao \ri$ in Fig.~\ref{energy_e1} for
both incident  polarization states. As $\beta$ decreases,
the energy density of the transmitted beam is concentrated at lower
values of $x$. Indeed, for both incident $s$- and $p$-polarized beams,  the
transmitted beam emerges from the moving slab at locations with $x <
0$ when  $\beta = -0.5$. This is in consonance with the orientation
of  the corresponding cycle--averaged Poynting vectors, as presented
in Fig.~\ref{pv_e1}. We note that the maximum amplitude of the
energy density of the transmitted beam decreases as $\beta$
decreases for the incident $s$-polarization state, but this is not the case for the incident
$p$-polarization state. Also, the maximum amplitude of the energy density
of the reflected beam increases as $\beta$ decreases for the incident $s$-polarization
state but decreases as $\beta$ decreases for the other
incident polarization state. Extrapolating from Fig.~\ref{energy_e1}, we estimate
that at $\beta \approx 0.2$ the energy density of the  transmitted
beam would be concentrated at $x = 3 \lambdao$ on the
face $z=L$ of the slab. The beam would be essentially undeflected by the moving slab
at this relative speed and thereby a degree of concealment of the slab may
be achieved \c{Hiding}.

\begin{figure}[!ht]
\centering \psfull \epsfig{file=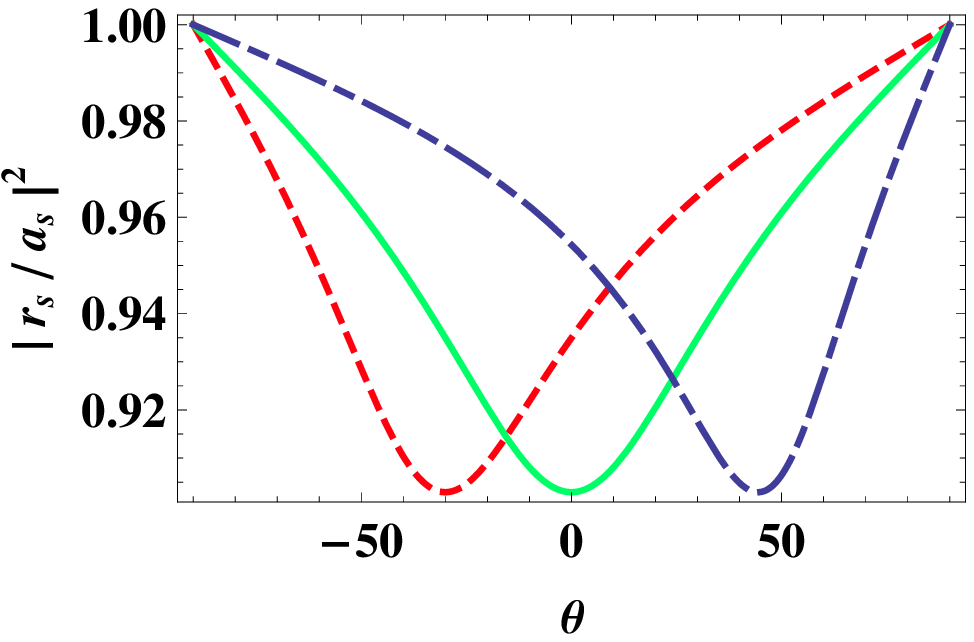,width=3.1in} \hfill
\epsfig{file=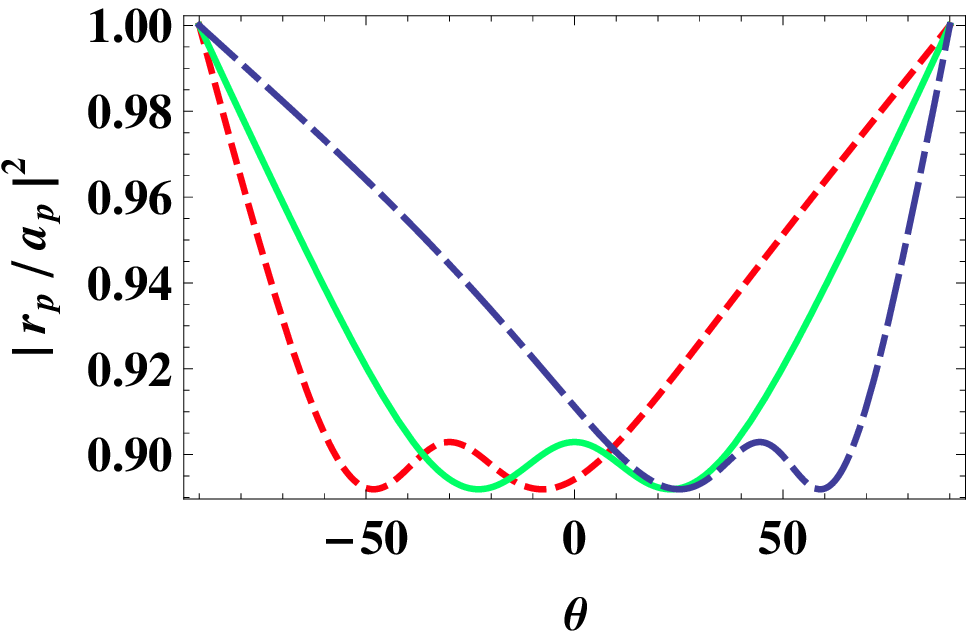,width=3.1in} \\
\epsfig{file=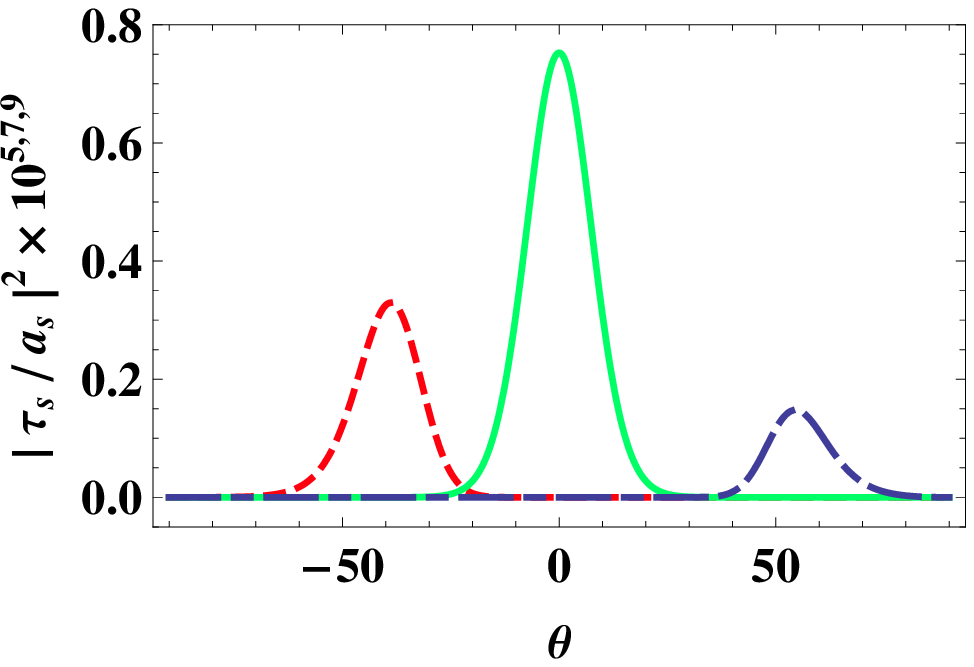,width=3.1in}  \hfill
 \epsfig{file=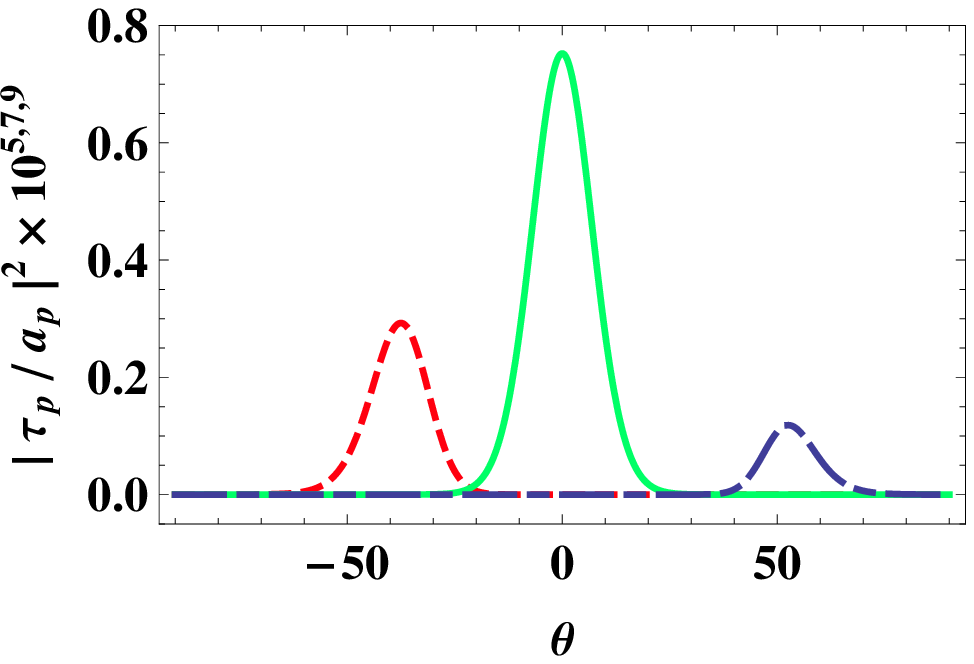,width=3.1in}
 \caption{\label{trans_e2}
The reflectances $\left| r_{u} / a_{u} \right|^2$ and transmittances
$\left| \tau_{u} / a_{u} \right|^2$,  $( u = s,p)$, plotted against
angle of incidence $\theta \in \le -90^\circ, 90^\circ \ri$,  for
$\beta = -0.5$   (broken curve,  red), $\beta = 0$ (solid curve,
green), and $\beta = 0.7 $ (broken dashed curve, blue). The
 moving slab of thickness $L  = 3 \lambdao$ is made of material B ($\eps' = - 0.34 + 0.04 i $).
The transmittances  are multiplied by $10^7$ for $\beta = -0.5$, by
$10^9$ for $\beta = 0$, and $10^5$ for $\beta = 0.7$. }
\end{figure}

Now, in view of \S\ref{num_1}, we turn to the case where the slab
is made of material B ($\eps' = -0.34 + 0.04i$). All other parameters are the same as those
used for Figs.~\ref{trans_e1} and \ref{energy_e1}.
 The reflectances
and transmittances are plotted
against the angle of incidence $\theta \in \le -90^\circ, 90^\circ
\ri$, for $\beta \in \lec -0.5, \, 0, \, 0.7 \ric$ in
Fig.~\ref{trans_e2}. The graphs of the  reflectances and
transmittances are symmetric about  $\theta = 0^\circ$  for $\beta =
0$, but they are asymmetric
 for $\beta \neq 0$.
We note that the magnitudes of the transmittances are very much
smaller than those of the reflectances, with the largest
transmittances occurring when the relative speed $\beta$ is largest.
 The oscillations apparent in Fig.~\ref{trans_e1} are largely absent
 from the graphs of the
reflectances
 and   transmittances in Fig.~\ref{trans_e2}. This is
 because of the much greater degree of attenuation within the uniformly moving slab
in the present case.

\begin{figure}[!ht]
\centering \psfull \epsfig{file=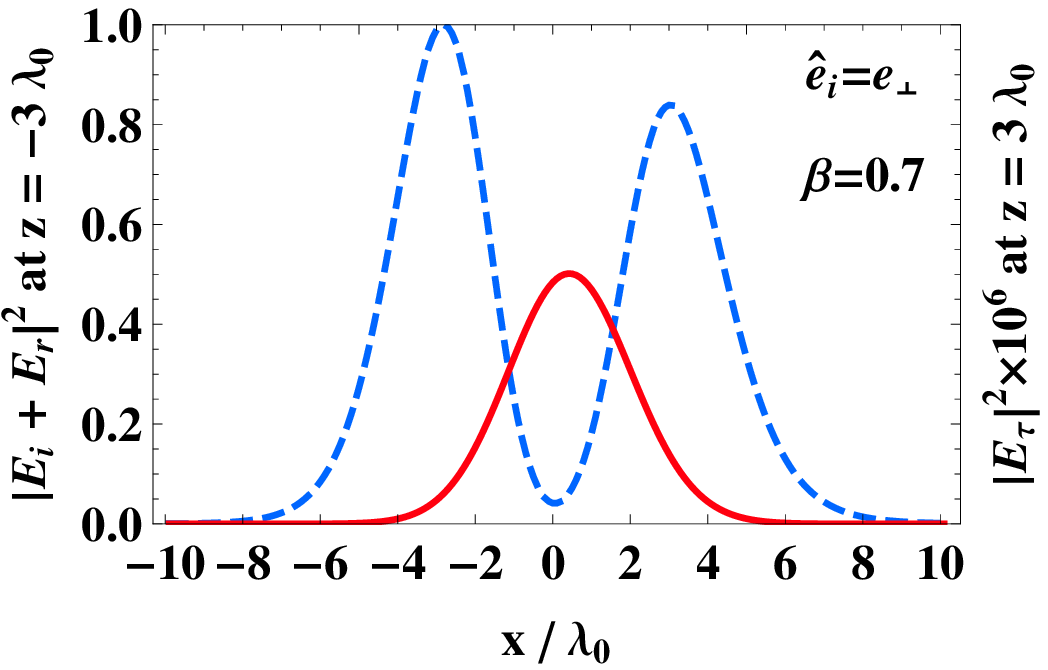,width=3.1in}
\\
\epsfig{file=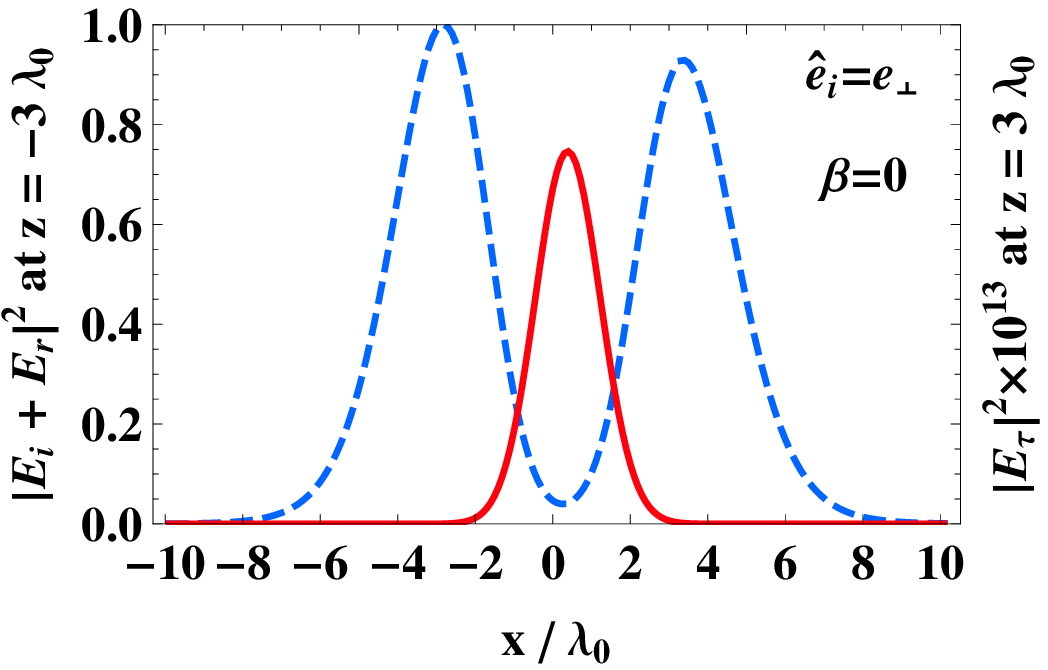,width=3.1in} 
  \\
\epsfig{file=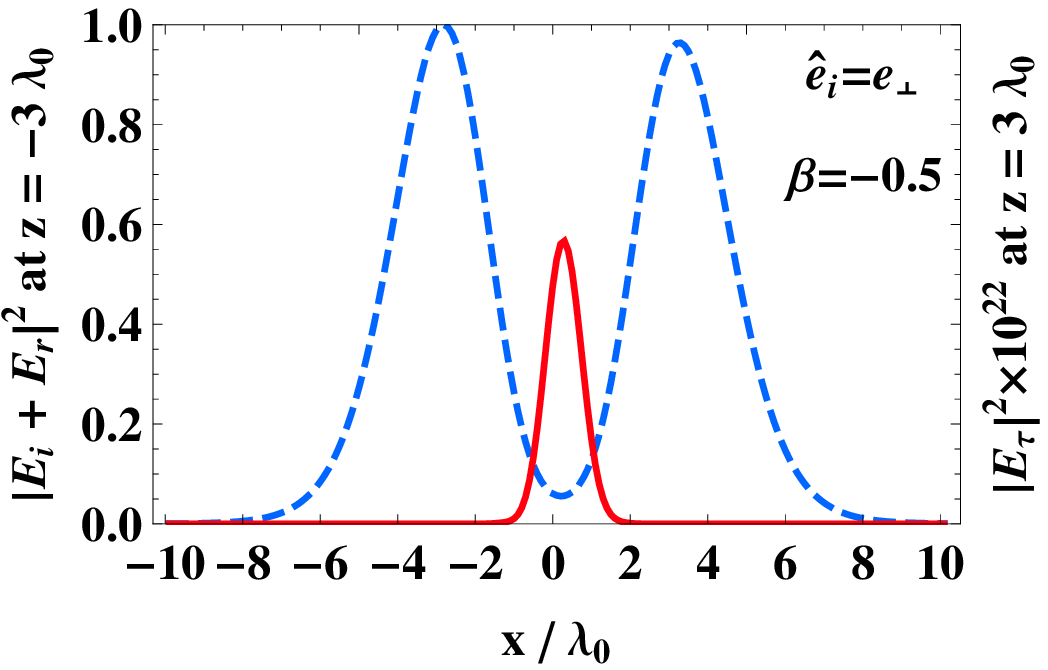,width=3.1in}  
 \caption{\label{energy_e2}
As Fig.~\ref{energy_e1} except that material A is replaced by
material B
 ($\eps' = - 0.34 + 0.04 i $). Only the graphs for the incident $s$--polarized beam are presented.
 }
\end{figure}

The beam energy densities at $z=-L$ and $z=L$ are mapped across $x
\in \le -10  \lambdao, \, 10 \lambdao \ri$ in Fig.~\ref{energy_e2}
for $\beta \in \lec -0.5, \, 0, \, 0.7 \ric$, at a mean angle of
incidence $\theta = 45^\circ$. Only the plots for the incident
$s$--polarized  beam are presented here; the corresponding plots for
the
 incident
$p$--polarized  beam are similar.
 For both incident
 polarization states, the energy density of the transmitted
beam~---~which is very much lower than that of the reflected
beam~---~decreases greatly as the relative speed $\beta$ decreases.
This is in keeping with the behaviours of the corresponding
reflectances and transmittances, as presented in
Fig.~\ref{trans_e2}. Furthermore, the energy density of the
transmitted beam is largely directed normally, away from  the
interface $z=L$. This may be explained by considering the direction of the
corresponding cycle--averaged Poynting vector, as presented in
Fig.~\ref{pv_e2}. The cycle--averaged Poynting vector is
approximately aligned with $\hat{\#z}$ for an angle of incidence of
approximately $55^\circ$ for $\beta = 0.7$. Thus, the maximums of
the transmittances at  $\theta \approx 55^\circ$ for $\beta = 0.7$
give rise to the  maximums in the  transmitted energy density in the
direction of $\hat{\#z}$ for $\beta = 0.7$  observed in
Fig.~\ref{energy_e2}. And similarly for the instances where  $\beta
= 0$ and $\beta = -0.5$, but here the energy densities of the
transmitted beams are much lower.

\section{Concluding remarks}

By directly implementing the Lorentz transformations of the electric
and magnetic fields, we have confirmed that counterposition and NPV
are not Lorentz-covariant phenomenons, in  the context of
dissipative mediums. This finding had previously been established
for nondissipative mediums using  the Minkowski constitutive
relations \c{Kong_PRB,Hiding}. We further demonstrated that the
lateral position of a Gaussian beam, upon traversal through a
uniformly moving slab, as  perceived by a nonco--moving observer,
could be controlled by varying the slab's velocity, when the
velocity is tangential to the interfaces and lies wholly in the
plane of incidence. Two  instances are especially noteworthy: (i) at
certain velocities, the transmitted beam can emerge from the slab
laterally displaced
 in the direction opposite to the direction of incident beam; and
(ii) at a unique slab velocity, the beam can emerge from the slab
with no lateral shift in its position, thereby achieving a degree of
concealment for the slab \c{Hiding}. For a uniformly moving slab
made of a  material which supports both positive and negative phase
velocity when at rest (material B in the previous sections), it was
seen that
 the transmittances are very
small and the energy density of the transmitted beam is largely
concentrated in the direction normal to the slab, regardless of the
inertial frame of reference.

\normalsize

\newpage

\newpage

\newpage

\newpage

\newpage

\newpage

\newpage

\newpage

\begin{thebibliography}{99}


\bibitem{SAR}
Ramakrishna S A 2005   Physics of negative refractive index
materials \emph{Rep. Prog. Phys.} {\bf 68} 449--521

\bibitem{Laeu}
 Lakhtakia A, McCall  M W and  Weiglhofer W S 2002
Brief overview of recent developments on negative phase-velocity
mediums (alias left-handed materials) \emph{Arch. Elektron.
\"Ubertr.} {\bf 56}   407--410

\bibitem{EJP}
 McCall M W,  Lakhtakia A and  Weiglhofer W S 2002 The negative index
of refraction demystified \emph{Eur. J. Phys.} {\bf 23} 353--359

\bibitem{Belov_MOTL}
Belov P A 2003 Backward waves and negative refraction in uniaxial
dielectrics with negative dielectric permittivity along the
anisotropy axis
 \emph{Microwave
  Opt. Technol. Lett.} {\bf 37} 259--263

\bibitem{ML_PRB}
Mackay T G and  Lakhtakia A 2009 (submitted for publication)
$\mathsf{http://arxiv.org/abs/0903.1530}$


\bibitem{Chen}
 Chen H C 1983 \emph{Theory of Electromagnetic
Waves} (New York: McGraw--Hill)


\bibitem{ZFM}
 Zhang Y, Fluegel B and  Mascarenhas A 2003
 Total negative refraction in real crystals for ballistic electrons and light
 \emph{ Phys. Rev. Lett.} {\bf 91}
157404

\bibitem{Optik_counterposition}
 Lakhtakia A and  McCall M W 2004 Counterposed phase velocity and
energy--transport velocity vectors in a dielectric--magnetic
uniaxial medium \emph{Optik} {\bf 115} 28--30

\bibitem{Kong_PRB}
 Grzegorczyk T M  and  Kong J A
2006 Electrodynamics of moving media inducing positive and negative
refraction \emph{Phys. Rev. B} {\bf 74} 033102


\bibitem{MOTL_counterposition}
 Mackay T G and  Lakhtakia A 2007
 Counterposition and negative refraction due to uniform motion
 \emph{ Microwave
  Opt. Technol. Lett.} {\bf 49} 874--876

\bibitem{note1}{We note that
 the phenomenon of counterposition is referred to as
`negative refraction' in Ref.~\cite{Kong_PRB}. In contrast, and in
keeping with Snel's law \c{Chen}, we adopt the standard convention
wherein the sense of refraction is
 determined solely by the relative orientations of the real parts of the
 refraction and incidence wavevectors.}

\bibitem{Hiding}
Mackay T G and   Lakhtakia A 2007
 Concealment by uniform motion
\emph{J.  Euro. Opt. Soc.~--~Rapid Pub.}
  {\bf 2}  07003

\bibitem{BC}
Besieris I M and   Compton Jr R T 1967 Time--dependent Green's
function for electromagnetic waves in moving conducting media
\emph{J. Math. Phys.} {\bf 8}  2445--2451


\bibitem{ML_Optik}
 Mackay T G  and  Lakhtakia A 2009
Positive--, negative--, and orthogonal--phase--velocity propagation
of electromagnetic plane waves in a simply moving medium:
reformulation and reappraisal \emph{Optik} {\bf 120} 45--48


\bibitem{Ditchburn}
 Ditchburn R W and  Freeman G H C 1966
 The optical constants of aluminium from 12 to 36 eV
\emph{Proc. R. Soc. Lond. A } {\bf 294} 20--37


\bibitem{Haus}
 Haus H A 1984  {\em Waves and Fields in Optoelectronics\/}
(Englewood Cliffs, NJ: Prentice--Hall)



\end{thebibliography}
\end{document}